\documentclass[amsmath,amssymb,aps,superscriptaddress,amsart]{revtex4-2}   

\usepackage{graphicx}
\usepackage[hidelinks]{hyperref}
\usepackage{xcolor}
\usepackage{booktabs}
\usepackage{multirow}
\usepackage{wasysym,MnSymbol}

\usepackage{ifsym,bbding}
\usepackage{amsmath}
\usepackage{slashbox}
\usepackage{footmisc}[symbol]

\makeatletter
\newcommand\footnoteref[1]{\protected@xdef\@thefnmark{\ref{#1}}\@footnotemark}
\makeatother

\begin{document}

\title{Reynolds-number effects on the outer region of adverse-pressure-gradient turbulent boundary layers}

\author{Rahul Deshpande{\footnote{\label{note1}both these authors contributed equally}}}
\email{raadeshpande@gmail.com}
\affiliation{Department of Mechanical Engineering, University of Melbourne, Parkville, VIC 3010, Australia}

\author{Aron van den Bogaard\footnoteref{note1}}
\affiliation{Department of Mechanical Engineering, University of Melbourne, Parkville, VIC 3010, Australia}
\affiliation{Physics of Fluids Group, University of Twente, P.O. Box 217, 7500AE Enschede, Netherlands}

\author{Ricardo Vinuesa}
\affiliation{FLOW, Engineering Mechanics, KTH Royal Institute of Technology, Stockholm, 10044, Sweden}%

\author{Luka Lindić}
\affiliation{Department of Mechanical Engineering, University of Melbourne, Parkville, VIC 3010, Australia}

\author{Ivan Marusic}
\affiliation{Department of Mechanical Engineering, University of Melbourne, Parkville, VIC 3010, Australia}

\begin{abstract}

We study the Reynolds-number effects on the outer region of moderate adverse-pressure-gradient (APG) turbulent boundary layers (TBLs) and find that their small scale (viscous) energy reduces with increasing friction Reynolds-number ($Re_{\tau}$).
The trend is based on analyzing APG TBL data across 600 $\lesssim$ $Re_{\tau}$ $\lesssim$ 7000, and contrasts with the negligible variation in small viscous-scaled energy noted for canonical wall flows.
The datasets considered include those from a well-resolved numerical simulation
[Pozuelo \emph{et al.},  {J.\ Fluid Mech.}, \textbf{939}, A34 (2022)],
which provides access to an APG TBL maintained at near-equilibrium conditions across 1000 $\lesssim$ $Re_{\tau}$ $\lesssim$ 2000, with a well-defined flow history, and a new high-$Re_{\tau}$ ($\sim$ 7000) experimental study from the large Melbourne wind tunnel, with its long test section modified to permit development of an APG TBL from a `canonical' upstream condition.
The decrease in small scale energy with $Re_{\tau}$ is revealed via decomposing the streamwise normal stresses into small and large scale contributions, based on a sharp spectral cut-off.
The origin for this trend is traced back to the production of turbulent kinetic energy in an APG TBL, the small scale contribution to which is also found to decrease with $Re_{\tau}$ in the outer region.
The conclusion is reaffirmed by investigating attenuation of streamwise normal stresses due to changing spatial-resolutions of the numerical grid/hotwire sensors, which reduces with increasing $Re_{\tau}$ and is found to be negligible at $Re_{\tau}$ $\sim$ 7000 in this study.
The results emphasize that new scaling arguments and spatial-resolution corrections should be tested rigorously across a broad $Re_{\tau}$-range, particularly for pressure gradient TBLs.

\end{abstract}

\maketitle

\section{Introduction}\label{sec:1intro}

Turbulent boundary layers (TBLs) have been a topic of active research for over a century, considering their relevance to many engineering and scientific applications. 
Despite such dedicated efforts, progress has been slow owing to the unique complexities (in the form of parametric dependencies) associated with each application.
While the Reynolds-number is the sole parameter influencing the flow physics for incompressible, zero-pressure-gradient (ZPG) and smooth-wall (\textit{i.e.} canonical) TBLs, wall-bounded flows found in nature are almost always exposed to non-canonical complexities, bringing dependence on additional parameters \citep{clauser1956turbulent,devenport2022}.
In the present study, we limit our focus to smooth-wall boundary layers exposed to an adverse-pressure-gradient (APG), which are commonly found over airplane wings and turbine blades, in diffusers, as well as many other applications.

Apart from the friction Reynolds-number ($Re_{\tau}$), previous studies have found the APG strength as well as its upstream-history to influence the smooth-wall APG TBL characteristics.
Following \citet{vinuesa2016}, here, $Re_{\tau}$ $=$ ${\delta_{99}}{U_\tau}/{\nu}$, where $\delta_{99}$ is the TBL thickness, $U_\tau$ is the friction velocity and $\nu$ is kinematic viscosity.
For a statistically two-dimensional mean flow, the strength of the pressure-gradient at a given streamwise location $x$ is typically quantified by the Clauser pressure-gradient parameter, ${\beta}(x)$ $=$ $(\delta^* / {\rho U_\tau^2}) \left( {\rm d} P/ {\rm d} x \right)$ \citep{clauser1954,clauser1956turbulent}, where $\delta^* = \int_0^{\delta_{99}}(1 - U(z)/U_e) {\rm d}z$ is the displacement thickness, $U(z)$ is the mean streamwise velocity at wall-normal location $z$, $U_e$ $=$ $U$($z$ = $\delta_{99}$), \textit{i.e.} the edge velocity, $\rho$ is the fluid density and ${\rm d} P/{\rm d} x$ is the mean streamwise pressure-gradient at $x$.
The influence of these additional parameters makes the APG TBL characteristics inconsistent with some popular scaling arguments based on canonical wall flows.
For example, the mean streamwise velocity profiles for an APG TBL have been reported to deviate from the classical log-law \citep{coles1956law,marusic2013,knopp2021experimental}.
The APG also reduces the wall-normal extent of the log-region, which is accompanied by the increased size of its wake region \citep{skaare1994turbulent,devenport2022}.
Most of these changes can be associated with modifications in the inherent structure of the turbulent flow on exposure to an APG \citep{lee2017,vinuesa2018,bross2019}.
These modifications are made evident by the wall-normal variation of the streamwise normal stress, which is given by the variance of the streamwise velocity fluctuations ($\overline{u^2}$).
In the case of ZPG TBLs, $\overline{u^2}$ profiles have the most prominent peak in the inner region of the TBL (referred henceforth as the inner peak), with the maximum $\overline{u^2}$ values in the outer region much lower than the inner peak at low-to-moderate $Re_{\tau}$ \citep{hutchins2009}.
In contrast, APG TBLs in a similar $Re_{\tau}$ range but with a reasonably strong $\beta$ will exhibit a distinct local maximum of $\overline{u^2}$ in the outer region (\textit{i.e.} an outer peak), which can be larger than the inner peak. \citep{skaare1994turbulent,harun2013,bobke2017history}.
This prominent outer peak is associated with enhanced contributions from the large as well as small scales, as compared to a ZPG TBL \citep{skaare1994turbulent,pozuelo2022,gungor2022}.
Such unique observations have motivated the proposal of new scaling arguments for the outer region of APG TBLs \citep{clauser1956turbulent,castillo2001,knopp2015,kitsios2016direct,maciel2018outer,gibis2019,pozuelo2023,wei2023} which, however, are based on a limited parameter space. 
What is ideally desired is a universal scaling that can account for the multi-parameter dependence \citep{devenport2022}, as comprehensively as that noted previously for canonical flows \citep{coles1956law,zagarola1998,marusic2013,chen2021}.

\begin{table}[t!]
\begin{center} 
\centering
\begin{tabular}{cccccc}
\hline\noalign{\smallskip}
\multicolumn{1}{c}{} & \multicolumn{2}{c}{ZPG TBL} & \multicolumn{3}{c}{APG TBL} \\
\cmidrule{2-2} \cmidrule{4-6}
\backslashbox{Scales}{Parameters} & $Re_{\tau}$ $\uparrow$ & & $Re_{\tau}$ $\uparrow$ & $\beta$ $\uparrow$ & $\overline{\beta}$ $\uparrow$ \\
\noalign{\smallskip}\hline\noalign{\smallskip}
Small scale   &  \textbf{Negligible}    &    & {\color{red}\textbf{Decreases?}}     & \textbf{Increases}        & \textbf{Increases}  \\
energy        &   \citet{hutchins2009}  &    & \emph{present}               &  \citet{monty2011}        & \citet{tanarro2020}   \\
              &                         &    & \emph{hypothesis}            & \citet{bobke2017history}  &                       \\
              &                         &    &                       &                           &                       \\ 
Large scale   &  \textbf{Increases}     &    & \textbf{Increases}    & \textbf{Increases}        & \textbf{Increases}  \\
energy        &  \citet{hutchins2009}   &    & \citet{tanarro2020}   & \citet{monty2011}         & \citet{tanarro2020}  \\
              &                         &    &                       & \citet{bobke2017history}  &                       \\
\noalign{\smallskip}\hline
\end{tabular}
\end{center}
\caption{Effects on the small (${\lambda}_{y}$ $\leq$ ${\lambda}_{y,c}$) and large scale (${\lambda}_{y}$ $>$ ${\lambda}_{y,c}$) energy in the outer region ($z^+$ = $z{U_{\tau}}/{\nu}$ $\gtrsim$ 100) of an APG TBL on independently increasing the three concerned parameters. 
It should be noted that this summary is applicable only for moderate APG ($\beta \ \lesssim 2$) and low-to-moderate $Re_{\tau}$ ($500 \leq $ Re$_\tau \leq 3000$) ranges investigated in the cited sources. 
In the case of ZPG TBL, however, the analysis is based on 3000 $\lesssim$ $Re_{\tau}$ $\lesssim$ 20000 \citep{hutchins2009}.
${\lambda}_{y,c}$ corresponds to a nominal spanwise wavelength cut-off to segregate small (fine dissipative and viscous scales) and large (inertia-dominated) scales.}
\label{tab0}
\end{table}

Establishment of scaling laws, inevitably, is preceded by a thorough understanding of the influence of various parameters on the flow characteristics.
This is the primary aim of the present study, which limits its focus to the outer region ($z^+$ = $z{U_{\tau}}/{\nu}$ $\gtrsim$ 100) of a smooth wall TBL exposed to a moderate APG ($\beta$ $\lesssim$ 2), a regime which has not been extensively studied, despite its importance \citep{romero2022properties}.
The present discussion is also limited to three parameters -- $Re_{\tau}$, $\beta$($x$) and upstream-history of $\beta$ (\textit{i.e.} $\overline{\beta}$($x$)) -- which have been investigated systematically by multiple past studies \citep{monty2011,bobke2017history,sanmiguel2017adverse,vinuesa2018,sanmiguel2020a,tanarro2020}.
Their findings have significantly advanced our understanding of the outer region, and have been summarized in table \ref{tab0} for convenience.
Here, $\overline{\beta}$($x$) is used to quantify upstream flow-history effects \citep{vinuesa2017,sanmiguel2020b} and is estimated by:
\begin{equation}
\label{eq1}
\begin{aligned}
{\overline{\beta}}(x) = \frac{\int_{\text{Re}_{\theta,x=0}} ^{\text{Re}_{\theta,x}} \beta (\text{Re}_{\theta,x}) {\:} d\text{Re}_{\theta,x}}{\text{Re}_{\theta,x} - \text{Re}_{\theta,x=0}}, 
\end{aligned}
\end{equation}
where $Re_{\theta}$ corresponds to the Reynolds-number based on momentum thickness (=${\theta}{U_e}/{\nu}$).
It is important to note here that an APG TBL can also be characterized by several other parameters \citep{coles1956law,townsend1976,castillo2001,perry2002,monty2011,harun2013,bobke2017history,pozuelo2022,devenport2022}, most of which have been historically used to quantify equilibrium/self-preserving aspects of these TBLs, which is beyond the scope (and focus) of this study.
Here, we discuss the effects of the three selected parameters on the small (${\lambda}_{y}$ $\leq$ ${\lambda}_{y,c}$) and large scale (${\lambda}_{y}$ $>$ ${\lambda}_{y,c}$) energy (table \ref{tab0}), where ${\lambda}_{y,c}$ corresponds to a representative spanwise wavelength cut-off for segregating these scales (with ${\lambda}_{y}$ = 2${\pi}$/$k_{y}$; $k_{y}$ being the spanwise wavenumber).
Throughout this manuscript, by small scales, we are essentially referring to the fine dissipative scales and the viscous-scale contribution to $\overline{u^2}$, while large scales correspond to the contribution from inertial scales (\textit{i.e.} attached eddies and superstructures; \citep{smits2011}).
Previous studies have found both the small and large scales to be energized with increase in $\beta$ as well as ${\overline{\beta}}$($x$) \citep{monty2011,bobke2017history,tanarro2020}.
Similarly, \citet{tanarro2020} have noted that an increase in $Re_{\tau}$ increases the large scale energy in APG TBL, similar to the behaviour noted for ZPG TBL \citep{hutchins2009}.
However, none of the previous studies have investigated the $Re_{\tau}$ effects on the small scale energy, leaving open an important question.
It is important because small scales from the inner region are advected away from the wall under a moderate APG \citep{tanarro2020}, substantially increasing their contribution to the turbulent kinetic energy in the outer region \citep{pozuelo2022,gungor2022,pozuelo2023}.
This preliminary analysis of \citet{tanarro2020}, however, was based on low-$Re_{\tau}$ simulations over an airfoil, which also suggested a decreasing influence of the APG with increasing $Re_{\tau}$.
This makes it imperative to understand the small scale dependency on $Re_{\tau}$, especially from the perspective of developing robust scaling arguments for the outer region.
A hypothesis that naturally follows from the above discussion is the reduction in small scale energy with $Re_{\tau}$.
However, the hypothesis is in contrast with the viscous-scaled energy of the small scales in ZPG TBLs, which is influenced very weakly by $Re_{\tau}$ in the outer region ($\sim$ $Re^{-1/4}_{\tau}$ \citep{chen2021}), making it quasi $Re_{\tau}$-invariant or negligible (table I).
This is the primary motivation of the present study, which investigates the $Re_{\tau}$-effects on the small scale energy in the outer region of an APG TBL, to potentially answer the open question highlighted in table \ref{tab0}.

Understanding this effect on the small scales is also important from the perspective of quantifying measurement errors.
As discussed, establishing and testing scaling arguments for an APG TBL would require high-$Re_{\tau}$ data, available typically through experiments \citep{knopp2015,romero2022properties,wei2023}.
These measurements are conducted either using multi-component anemometry \citep{marusic1995,deshpande2020,romero2022properties}, particle image velocimetry \citep{knopp2015,cuvier2017,knopp2021experimental} or laser-doppler velocimetry \citep{aubertine2005,volino2020non}.
However, data from all of these measurement techniques inevitably suffers from spatial-resolution issues, leading to attenuated estimates of the small scale energy \citep{ligrani1987,hutchins2009}.
This spatial-resolution issue, which is typically known to bring inaccuracies in the near-wall region, can however, also adversely affect measurements in the outer region of an APG TBL, where small scales are statistically significant (at least at low-$Re_{\tau}$).
This can be detrimental to high $Re_{\tau}$ testing of scaling arguments, proposed based on low $Re_{\tau}$ data (or vice versa).
Although there are certain correction schemes to account for the spatial-resolution issues \citep{monkewitz2010,chin2011,smits2011spatial,jhlee2016}, they depend on direct numerical simulation (DNS) datasets/concepts based on canonical wall-bounded flows, which are not applicable for APG TBLs \citep{deshpande2020,sanmiguel2020a}.

To this end, the present study investigates moderate APG TBLs on a smooth flat wall across 600 $\lesssim$ $Re_{\tau}$ $\lesssim$ 7000.
It uses the published, well-resolved large-eddy simulation (LES) dataset of \citet{pozuelo2022} across 600 $\lesssim$ $Re_{\tau}$ $\lesssim$ 2000, and a new experimental APG TBL dataset from the large Melbourne wind tunnel, at $Re_{\tau}$ $\sim$ 7000.
Both these datasets correspond nominally to the same $\beta$ range, 1 $\lesssim$ $\beta$ $\lesssim$ 2, thereby permitting analysis across a decade of $Re_{\tau}$.
The present investigation is conducted through two approaches -- spectral decomposition methodologies ($\S$\ref{sec:4Re_effect}) as well as estimating velocity statistics for different numerical/experimental spatial-resolutions ($\S$\ref{sec:5spat_res}).
Consistent conclusions are obtained from both approaches, which confirms that our results are not a function of the sharp spectral cut-off (${\lambda}_{y,c}$) used for the first approach.
Throughout this manuscript, fluid properties with superscript `$+$' represent normalization by viscous velocity ($U_{\tau}$), length (${\nu}/{U_{\tau}}$) and time (${\nu}/{U^{2}_{\tau}}$) scales.
Flow properties in capital, or with overbars, represent mean spatial-/time-averaged properties while those in lowercase represent fluctuations about their respectives means.
Cartesian co-ordinates $x$, $y$ and $z$ represent the streamwise, spanwise and wall-normal directions, respectively, with $u$, $v$ and $w$ representing the associated velocity components.

\section{Experimental setup and numerical database}\label{sec:2setup}

This section gives details on the experimental and numerical data analyzed in the study.
Both these datasets are unique with regards to the manner in which the experiments/simulations were set up, enabling growth of an APG TBL with a well-defined upstream flow history. 
More details are provided in the sub-sections below.

\subsection{Experimental setup}\label{ss:2exp}
\begin{figure}[t]
    \centering
    \includegraphics[width=1.0\textwidth]{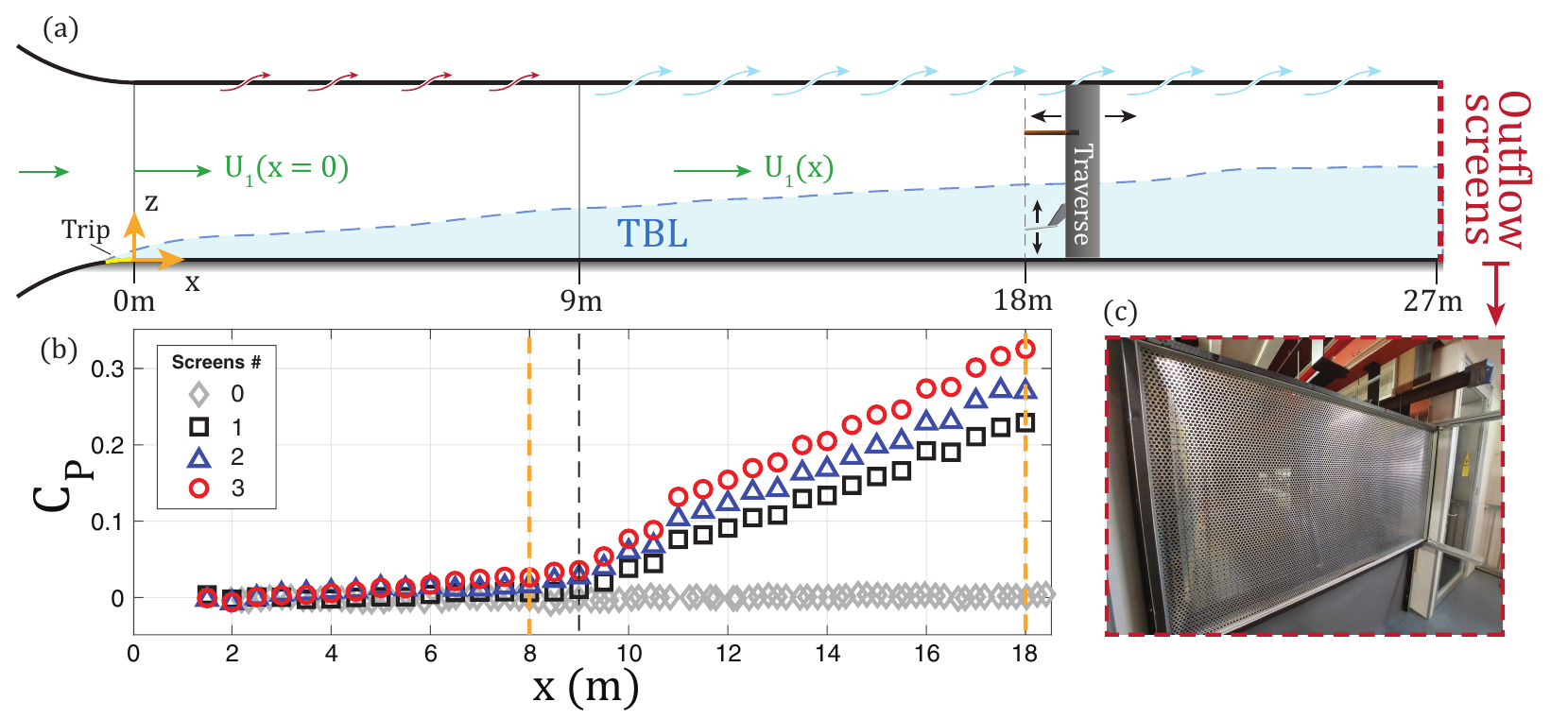}
    \caption{(a) Schematic overview of the large Melbourne wind-tunnel test-section modified to set up an APG TBL (not to scale). 
    Markings in red indicate the modifications made for the APG set up, which include (i) controlling the air bleeding from the ceiling in the upstream section ($x$ $<$ 9\:$m$) to enforce a nominal ZPG, and (ii) introducing outflow screens at the test-section exit to increase tunnel static pressure. 
    (b) Mean pressure coefficient, $C_P$ measured at various streamwise locations, $x$, for different Screen\# configurations -- 0, 1, 2 or 3, with 0 corresponding to the scenario when ZPG is maintained across the entire tunnel test-section.  
     (c) Photograph of an aluminium outflow screen installed at the test-section outlet. The screen had circular holes of 9.5\:$mm$ diameter at a pitch of 12.7\:$mm$, resulting in a porosity of 51\%.
    Dashed grey line in (b) refers to the $x$ location up to which a nominal ZPG was maintained, while dashed golden lines indicate locations for hotwire measurements. }
    \label{fig:2_1tunnelschematic}
\end{figure}

Experiments were conducted in the open-circuit, large Melbourne wind-tunnel at the University of Melbourne.
Its test-section has a cross-section of $\simeq$ 1.89 $\times$ 0.92\:$m^2$ and a long working section length of 27\:$m$, which permits development of a thick turbulent boundary layer ($\gtrsim$ 0.3\:$m$) towards its downstream end.
This aspect, combined with the ability to have free-stream speeds up to $\sim$ 40\:$m{s^{-1}}$, enables investigation of high-$Re_{\tau}$ TBLs in this facility.
Prior to this study, the wind-tunnel was predominantly used for ZPG TBL research, and details regarding characterization of the associated set-up can be found in \citet{marusic2015}.
In this facility, a ZPG TBL is set-up by installing a high porosity screen/mesh at the test-section outlet, which builds up a nominal back-pressure in the test-section.
The ceiling of the test-section comprises multiple air bleeds, spanning across the tunnel width and distributed at regular intervals along $x$ ($\sim$ 1.2\:$m$; figure \ref{fig:2_1tunnelschematic}a), which allow excess air to escape owing to pressure build up. 
At the same time, it also removes the TBL developing on the ceiling.
Such an arrangement establishes a controlled ZPG across the entire test-section length \citep{marusic2015}.
In the present manuscript, the high porosity screen used for the ZPG TBL is henceforth referred to as the Screen\#0 configuration, which acts as the base configuration for all the APG experiments.

Setting up an APG TBL at high-$Re_{\tau}$, with minimum upstream-history effects, demands unique modifications to the test-section.
The present study adopts the methodology used in the pioneering study of \citet{clauser1954}, where a gradual streamwise increment in the test-section static pressure was achieved by introducing blockage at the outlet section.
Figure \ref{fig:2_1tunnelschematic}(c) shows a photograph of an aluminium screen, having a low porosity of 51\%, installed at the outlet of the tunnel test-section as a blockage.
The screen is held rigidly on an aluminium frame at the tunnel outlet, with capacity to hold as many as six screens in total.
In the present study, we either fix 1, 2 or 3 of these low porosity screens in the frame (referred henceforth as configuration \#1, 2 or 3), to systematically increase the static pressure in the test-section.
In the upstream part of the test-section, however, a nominal ZPG is maintained always by controlling the outflow from the air bleeds on the ceiling, up to $x$ $\lesssim$ 9\:$m$ (indicated by red arrows in figure \ref{fig:2_1tunnelschematic}a).
This ensures that the APG TBL growing beyond $x$ $>$ 9\:$m$ begins with a high-$Re_{\tau}$ `canonical' upstream condition, \textit{i.e.} with negligible upstream-history effects.
Dynamic pressure (${\rho}{U^2_{1}}(x)$/2) was measured at various streamwise locations using a Pitot tube attached to a ($x,z$) traverse (figure \ref{fig:2_1tunnelschematic}a; \citet{marusic2015}) at $z$ = 0.525\:$m$ from the wall, which is always in the free-stream of both the top- and bottom-wall boundary layers.
The Pitot tube is connected to a 10 Torr Baratron manometer and signal conditioner.
Atmospheric pressure and temperature were also acquired during the experiment, which facilitate conversion of the dynamic pressure to free-stream speeds. 
The resulting pressure coefficient, $C_P = 1 - (U_1(x) / U_1({x} = 0))^2$, where $U_1(x)$ is the free-stream velocity at $x$, is obtained for different Screen \# configurations and is shown in figure \ref{fig:2_1tunnelschematic}(b).
The plots confirm that a nominal ZPG condition was maintained for $x$ $<$ 9\:$m$, followed by a quasi linear growth in static pressure corresponding to the streamwise APG ($x$ $\gtrsim$ 9\:$m$).

Notably, the present set up adopted to impose streamwise pressure-gradients is different from those adapted by past experimental studies \citep{aubertine2005,cuvier2017,sanmiguel2020b,volino2020non,knopp2021experimental,romero2022properties}, which predominantly relied on working section inserts (either on the ceiling or the `wall') to establish the APG.
Inclusion of working section inserts, however, introduces a constriction, which initially imposes a favourable pressure-gradient, followed by a limited straight section for a ZPG, before finally leading to the APG region. 
Without due precaution, such setups can impose unique upstream histories that influence the scaling of the mean flow statistics of the APG flow \citep{bobke2017history,pozuelo2022}. 
To avoid these history effects, other studies have used an adjustable ceiling in the tunnel test-section \citep{skaare1994turbulent,marusic1995,monty2011,harun2013,vinuesa2014,sanmiguel2020b,drozdz2021,romero2022properties,volino2020non}, the height of which is adjusted to decelerate the incoming flow, and consequently generate a streamwise APG.
Implementing this, however, requires a modular test-section along with significant setup time, making it impractical for the large Melbourne wind-tunnel.

\begin{table}[t]  
\centering
\begin{center} 
\begin{tabular}
{cccccccccccccccc}
\hline\noalign{\smallskip}
Ref. & {\rotatebox[origin=c]{90}{Screens\#}} & ${U_{1}}(x = 0)$ & $x$ & ${U_{1}}(x)$ & ${\beta}(x)$ & Re$_{\tau}$ & Re$_{\theta}$ & $U_{\tau}$ & ${\delta}_{99}$ & ${\delta}^{*}$ & $t^+$ & {\rotatebox[origin=c]{90}{$T{U_{1}(x)}/{{\delta}_{99}}$}} & $l^+$ & Symbol & Source \\
 &  & ($m/s$) & ($m$) & ($m/s$) &  &  &  & ($m/s$) & ($m$) & ($mm$) &  &  & & & \\    
\noalign{\smallskip}\hline\noalign{\smallskip}
\textbf{--}  &  0  & \underline{11.9} &  \underline{8}  & \underline{11.9}  & \underline{0.0} & \underline{4150} & 10400 & 0.42 & 0.15 & 18 &  0.49 & 15200 & \textbf{22} & {\color{gray}$\largestar$} & \citet{hutchins2009} \\
\textbf{--}  &  3  & \underline{12.1} &  \underline{8}   &  \underline{12}    & \underline{0.0} & \underline{4100} & 12400 & 0.42 & 0.15 & 22 &  0.24 & 16100 & \textbf{14} & {\color{red}$\varhexstar$} & present \\
Exp-ZPG  &  0  & 10.3 & 21  & 10.3  & 0.0 & 5600 & 19400 & 0.33 & 0.27 & 38 &  0.34 & 17400 & \textbf{11} & {\color{gray}$\largediamond$} & \citet{hutchins2009} \\
Exp-APG  &  1  & 11.3 & 18  & 10.0  & 0.9 & 6500 & 27400 & 0.31 & 0.33 & 61 &  0.15 & 21700 & \textbf{10},60 & $\largesquare$ & present \\
Exp-APG  &  2  & 11.7 & 18  & 10.0  & 1.3 & 6750 & 30800 & 0.30 & 0.35 & 69 &  0.15 & 20400 & \textbf{10},60 & {\color{blue}$\largetriangleup$} & present \\
Exp-APG  &  3  & 12.1 & 18  & 10.0  & 1.7 & 6900 & 31500 & 0.29 & 0.37 & 73 &  0.15 & 19800 & \textbf{10},20,40,60 & {\color{red}$\largecircle$} & present \\
\noalign{\smallskip}\hline
\end{tabular}
\end{center}
\caption{Table summarizing the parameters associated with various hotwire measurements reported in this study. 
Definitions/terminologies have been provided either in $\S$\ref{sec:1intro} or \ref{sec:2setup}.
$Re_{\theta}$ is based on momentum thickness, $\theta = \int_0^{\delta_{99}}(1 - U(z)/U_e)(U(z)/U_e){\rm d} z$.
Numbers in bold correspond to hotwires with the best spatial-resolution, while underlined numbers represent cases with matched experimental conditions -- $x$, $U_1$, $\beta$ and $Re_{\tau}$ -- but different outflow screen configurations.}
\label{tab1}
\end{table}

All the data in figure \ref{fig:2_1tunnelschematic}(b) was obtained for free-stream conditions, $U_1$($x$ $=$ 18\:$m$) $\approx$ 10\:$m{s^{-1}}$.
This matches the tunnel conditions for all (except one) new hotwire measurements presented in this study, which were conducted at $x$ $\approx$ 18\:$m$ for Screen \#1, \#2 or \#3 configurations (\textit{i.e.} $\beta$ $>$ 0; table \ref{tab1}).
As highlighted in the table, these new hotwire measurements will be referred to as `Exp-APG' in the remainder of the manuscript, statistics from which will be compared with the published `Exp-ZPG' measurement ($\beta$ = 0) at nominally the same $Re_{\tau}$.
Apart from these main measurements, another new hotwire measurement was conducted relatively upstream at $x$ $\approx$ 8\:$m$ (with Screen\#3 configuration), \textit{i.e.} towards the end of the controlled ZPG region, to demonstrate consistency with previously-published ZPG TBL measurements at matched conditions (with Screen\#0 configuration; table \ref{tab1}).
The consistency between these two measurements has been demonstrated in appendix 1 and confirms the `canonical' upstream condition for the new Exp-APG experiments presented here.
The hotwire measurements across the entire TBL were also made possible via the built-in traverse system, which has a Renishaw optical encoder (resolution $\sim$ 0.1\:$\mu$m) to precisely record the relative wall-normal movement of the sensor.
Velocity statistics reported here were acquired across the $z$-range: 0.5\:$mm$ $\lesssim$ $z$ $\lesssim$ 0.525\:$m$, at logarithmically-spaced locations.
In addition to these full TBL profiles, hotwires of the best spatial-resolution (highlighted in bold in table \ref{tab1}) were also used to measure mean streamwise velocity very close to the wall, in and above the viscous sublayer (0.2\:$mm$ $<$ $z$ $\lesssim$ 0.5\:$mm$, \textit{i.e.} 3 $<$ $z^+$ $\lesssim$ 8).
Such a measurement was possible at these high-$Re_{\tau}$ owing to the physically thick TBL and reasonably large viscous length scale in the Melbourne wind-tunnel.
The near-wall positioning of the hotwires, at the start of the measurement, was undertaken by using a travelling microscope (to measure near-wall distance) and a multimeter (to track conduction effects to wall), with the latter limiting its location to $\gtrsim$ 0.2\:$mm$ above the wall.
$U$($z$) measured in the viscous sublayer was used to estimate mean skin-friction velocity ($U_{\tau}$) for both the Exp-ZPG and Exp-APG measurements, by forcing the measurements to fit over viscous-scaled mean velocity profiles ($U^+$ = $U/{U_{\tau}}$) from published datasets (figures \ref{fig:3_1meanuvar}a,c).
We note here that while there are uncertainties associated with this method of estimating $U_{\tau}$, the key conclusions from the Exp-APG dataset ($\S$\ref{sec:5spat_res}) are not dependent on the estimation method for $U_{\tau}$.

For the present study, the hotwire sensors were made in-house using Wollaston wire, soldered onto Dantec 55P05 or 55P15 probes depending on the diameter ($d$) and length ($l$) of the Platinum sensors.
This capability made it possible to conduct experiments with systematically varying spatial-resolution (10 $\lesssim$ $l^+$ = $l{U_{\tau}}/{\nu}$ $\lesssim$ 60) of the sensors.
Hotwire sensors of various lengths ($l$ $=$ 0.5, 1, 2 and 3\:$mm$) were fabricated by carefully etching the Wollaston wire with nitric acid \citep{perry1982}, to expose the Platinum sensor.
The sensors were made out of wires of either $d$ = 2.5\:$\mu$m or 5\:$\mu$m to maintain the aspect ratio, $l/d$ $\gtrsim$ 200, to avoid end-conduction effects \citep{hutchins2009}.
The measurements were made using an in-house Melbourne University Constant Temperature Anemometer (MUCTA), at an overheat ratio of 1.8.
Hotwire data was sampled during measurement at a frequency rate ($f_s$), and for total acquisition time ($T$), that satisfy $t^+ = U_\tau^2 / (f_s \nu) < 3$ and $TU_1(x) / \delta_{99} > 20000$, to have good temporal resolution and achieve statistical convergence, respectively \citep{hutchins2009}.
Since it was challenging to maintain steady free-stream speeds below 4\:$m{s^{-1}}$ in the wind-tunnel test-section, the hotwire calibration was conducted using two different arrangements, both in an in situ manner.
A low-speed calibration was conducted using an in-house jet calibrator (0--5\:$m{s^{-1}}$; \citep{xia2022}) and 0.1 Torr Baratron manometer, which enables precise control across this low-speed range.
Tunnel free-stream speeds (4--12\:$m{s^{-1}}$) were used to calibrate the sensor across this higher speed range by maintaining the sensor at the same ($x$,$z$) location, as the Pitot tube acquiring the reference free-stream speed.
Pre- and post-calibrations were performed for all hotwire measurements, in conjunction with the `intermediate single point recalibration' (ISPR;\citep{talluru2014}) method of traversing the hotwires to the free-stream, to correct for the hotwire sensor drift over the duration of measurement. 
Finally, all the sampled data was low-pass filtered at half the sampling frequency to avoid aliasing and high-frequency noise. 
All new measurements reported in this study (table \ref{tab1}) were conducted at least twice to confirm repeatability.

\begin{figure}[t]
    \centering
    \includegraphics[width=0.6\textwidth]{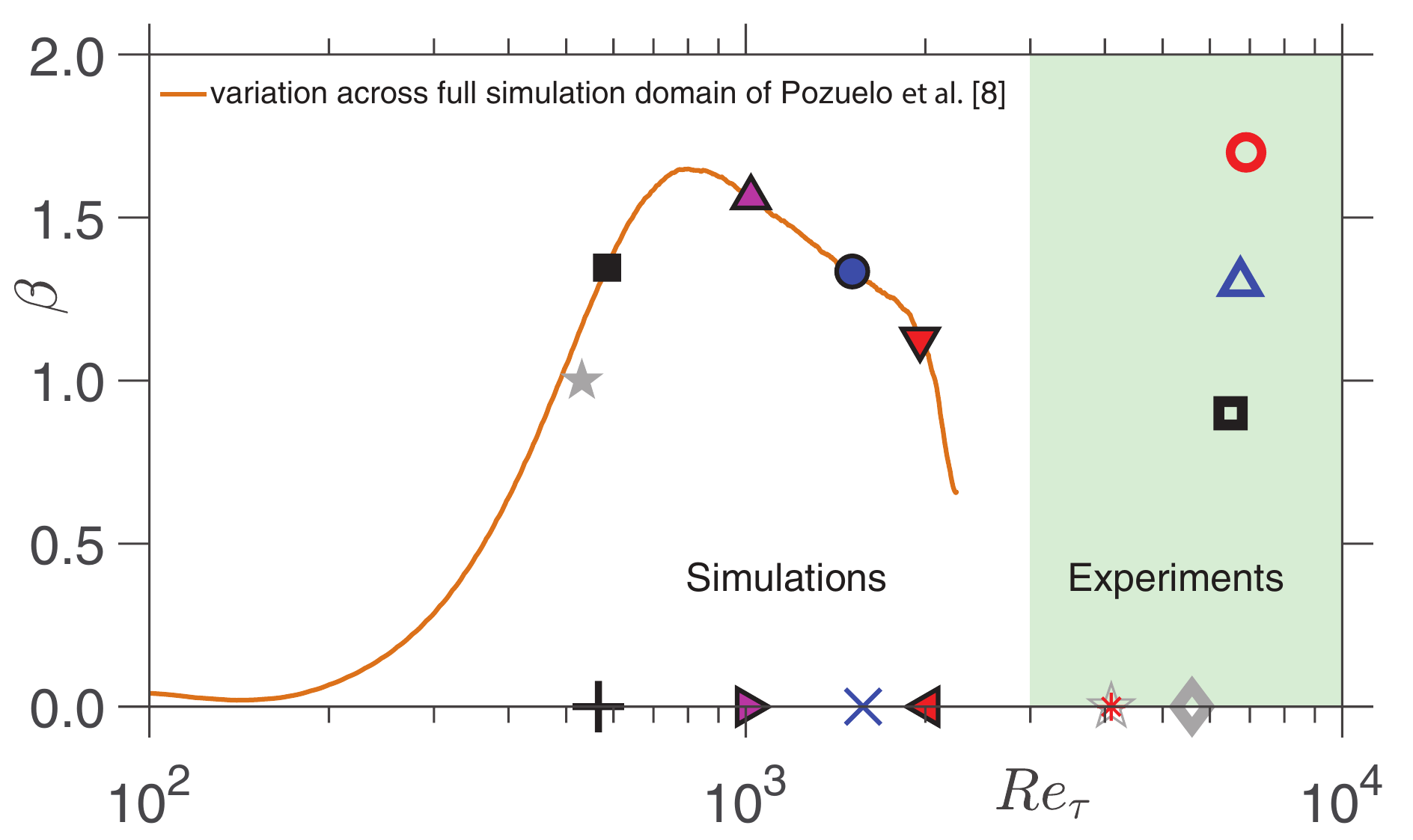}
    \caption{Parameter space $\beta$ vs Re$_\tau$ investigated in the present study using a combination of previously-published data \citep{hutchins2009,eitel2014simulation,bobke2017history,pozuelo2022} and new experimental data. 
    See tables \ref{tab1} and \ref{tab2} for information on respective datasets. 
    $\beta$ vs Re$_\tau$ is plotted across the full domain of the LES-APG dataset \citep{pozuelo2022} to indicate the upstream-history for the selected locations (given by symbols).}
    \label{fig:2_2betavsre}
\end{figure}

\begin{table}[t]  
\centering
\begin{center} 
\begin{tabular}
{ccccccccc}
\hline\noalign{\smallskip}
Ref. & $x$/${{\delta}^{*}_{o}}$ & ${\beta}(x)$ & ${\overline{\beta}}(x)$ & Re$_{\tau}$ & Re$_{\theta}$ & ${\Delta}{y^+}$ & Symbol & Sources \\   
\noalign{\smallskip}\hline\noalign{\smallskip}
LES-ZPG  &  1500  &  0   &  0    &  566  & 1638 & \textbf{13}          & \Plus & \citet{eitel2014simulation} \\
LES-ZPG  &  3500  &  0   &  0    & 1007  & 3136 & \textbf{12}          & {\color{magenta}$\filledmedtriangleright$} & \citet{eitel2014simulation} \\
LES-ZPG  &  6500  &   0   &   0    & 1572  & 5115 & \textbf{12}          & {\color{blue}$\times$} & \citet{eitel2014simulation} \\
LES-ZPG  &  9000  &   0   &   0    & 2017  & 6657 & \textbf{12}          & {\color{red}$\filledmedtriangleleft$} & \citet{eitel2014simulation} \\
\textbf{--}  &  1277  &   \underline{1.0} &   0.84 &  \underline{531}  & 2153 & \textbf{8} &   {\color{gray}$\filledlargestar$} & \citet{bobke2017history} \\
LES-APG  &  1500  &   \underline{1.3} &   0.67 &  \underline{585}  & 2362 & \textbf{11},21,43,64 & {\color{black}$\filledmedsquare$} & \citet{pozuelo2022} \\
LES-APG  &  3500  &   1.6 &   1.17 & 1020  & 4915 & \textbf{8},23,38,69  & {\color{magenta}$\filledmedtriangleup$} & \citet{pozuelo2022}  \\
LES-APG  &  6000  &   1.3 &   1.26 & 1508  & 7160 & \textbf{7},20,40,67  & {\color{blue}$\bullet$} & \citet{pozuelo2022}  \\
LES-APG  &  8500  &   1.1 &   1.26 & 1959  & 8959 & \textbf{6},19,44,63  & {\color{red}$\filledmedtriangledown$} & \citet{pozuelo2022}  \\
\noalign{\smallskip}\hline
\end{tabular}
\end{center}
\caption{Table summarizing the parameters associated with various numerical databases analyzed in this study. 
Definitions/terminologies have been provided either in $\S$\ref{sec:1intro} or \ref{sec:2setup}.
${\Delta}y^+$ corresponds to the various spanwise grid resolutions used to estimate the velocity statistics, with the numbers in bold corresponding to the original/well-resolved grid of the dataset.
Underlined numbers represent cases with reasonably similar $\beta$ and $Re_{\tau}$ but different upstream-history, $\overline{\beta}$($x$).}
\label{tab2}
\end{table}

The key TBL characteristics ($\beta$, $Re_{\tau}$) associated with all the hotwire experiments are documented in table \ref{tab1} and plotted in figure \ref{fig:2_2betavsre}.
These values have been reported based on the measurements conducted using hotwires with the best spatial-resolution ($l^+$ $\sim$ 10).
It should be noted that $Re_{\tau}$ in this study is estimated based on ${\delta}_{99}$, which was obtained via the diagnostic-plot method recommended by \citet{vinuesa2016}, for an APG TBL.
Accordingly, the $Re_{\tau}$ values associated with the measurements adopted from \citet{hutchins2009} have been recomputed based on the present definition.
Quantification of the Clauser pressure-gradient parameter ($\beta$) was made possible by combination of three different measurements: streamwise variation of test-section pressure to estimate ${\rm d}P/{\rm d}x$ (Pitot tube), velocity profile across the TBL to compute $\delta^*$ (hotwire), and near-wall measurements to estimate $U_{\tau}$ (hotwire) -- where \ ${\rm d} P/{\rm d}x$ was estimated based on a linear fit for $x$ $>$ 9\:$m$, in figure \ref{fig:2_1tunnelschematic}(b).
Since estimation of ${\overline{\beta}}$($x$) is challenging in the case of experiments, quantifying upstream-history of the Exp-APG dataset is beyond the scope of the present study.

\subsection{Numerical dataset}\label{ss:2les}

The present study also considers previously-published, well-resolved LES data of an APG TBL from \citet{pozuelo2022}.
This APG database is unique as it has a long computational domain, a significant portion of which is maintained at a nominally constant $\beta$, \textit{i.e.} at near-equilibrium condition.
This makes it one of the ideal datasets to investigate $ Re_\tau$-effects on an APG TBL, especially also considering its broad $Re_{\tau}$ range. 
For the purposes of referencing, this dataset will be referred to as the `LES-APG' dataset in the remainder of this manuscript.
Figure \ref{fig:2_2betavsre} shows the variation of $\beta$ with $Re_{\tau}$ across the full numerical domain, which is compared against the parameters associated with the present experiments (table \ref{tab1}).
Here, we also compare and contrast the observations from LES-APG with a well-resolved LES dataset of a ZPG TBL (henceforth referred as `LES-ZPG' \citep{eitel2014simulation}), which is available for the matched $Re_{\tau}$ range as the LES-APG.
For comparison, velocity statistics are extracted at four locations of the computational domain for both LES-APG and LES-ZPG, flow parameters for which have been highlighted with respective symbols in figure \ref{fig:2_2betavsre}.
In the case of LES-APG, the three locations at the highest $Re_{\tau}$ are associated with the near constant $\beta$ domain, while the lowest-$Re_{\tau}$ location is at similar $\beta$ range but upstream of the constant $\beta$ domain. 
Key TBL characteristics associated with these datasets are presented in table \ref{tab2}, with ${\Delta}y^+$ numbers in bold representing the original grid resolution of the datasets in the spanwise direction.
In order to investigate the influence of spatial-resolution using the numerical datasets, the velocity time-series data saved at the four $x$-locations was box-filtered \citep{chin2011} across multiple grid points along the span, to estimate statistics for 6 $\lesssim$ ${\Delta}y^+$ $\lesssim$ 69, consistent with 10 $\lesssim$ $l^+$ $\lesssim$ 60 range for the hotwires (table \ref{tab1}).

\begin{figure}[t]
    \centering
    \includegraphics[width=1.0\textwidth]{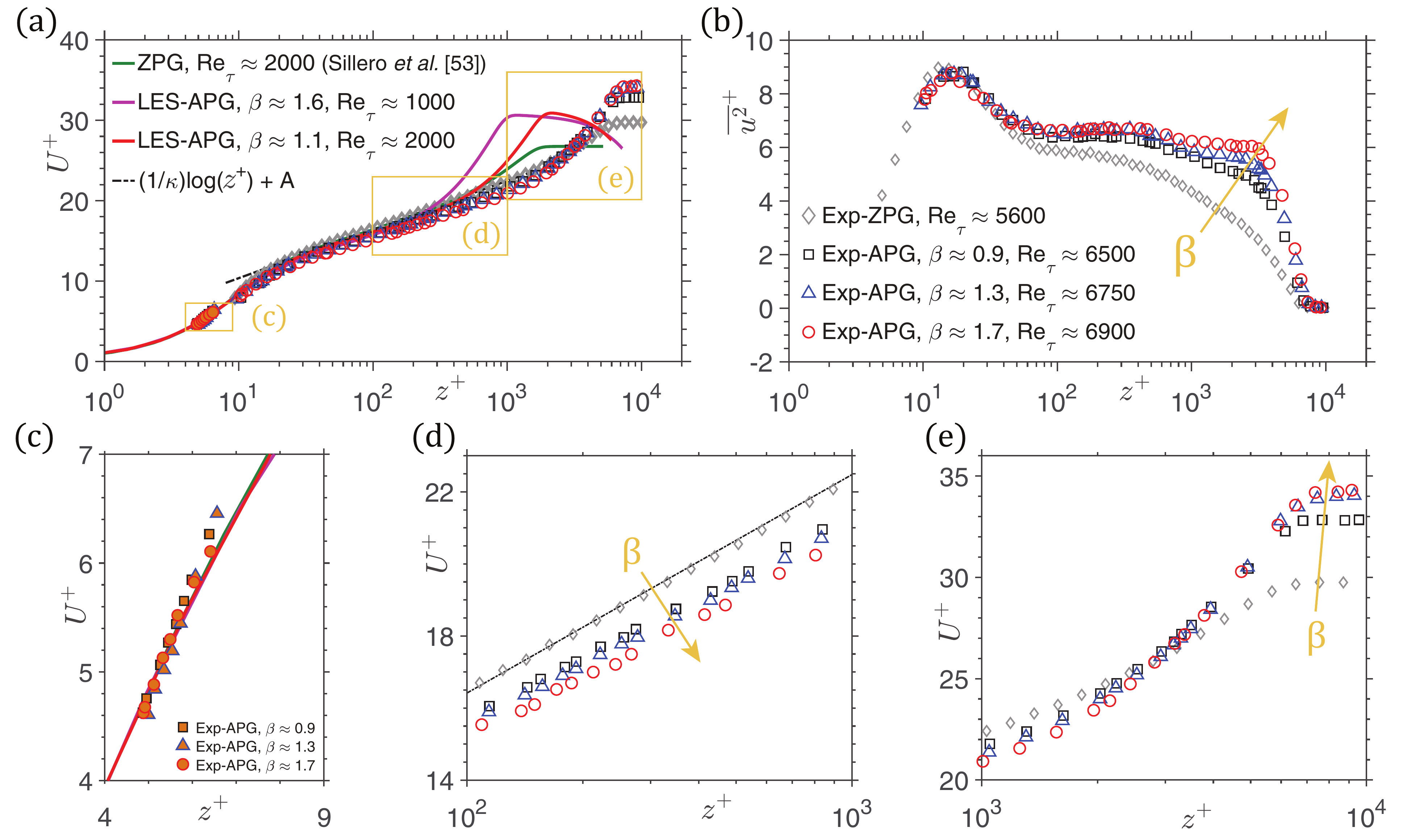}
    \caption{Viscous-scaled (a,c,d,e) mean velocity profiles and (b) streamwise normal stresses corresponding to the Exp-ZPG and Exp-APG datasets with $l^+$ $\approx$ 10 (table \ref{tab1}). 
In (a), the high-$Re_{\tau}$ experimental data is compared with low-$Re_{\tau}$ ZPG TBL dataset of \citet{sillero2014} and the well-resolved LES-APG dataset (table \ref{tab2}).
Plots in (c-e) depict the enlarged views of data in specific $z^+$ ranges in (a), indicated by yellow boxes.
For this figure, solid lines correspond to complete velocity profiles from simulation datasets, while open symbols correspond to experimental data acquired across the TBL.
Filled symbols in (a,c) are from the dedicated experiments conducted very close to the wall to estimate skin-friction velocity, $U_{\tau}$.
Dash-dotted black line in (a,d) corresponds to the well-known log-law of the mean velocity following \citet{marusic2013}, with $\kappa$ $=$ 0.39 and A = 4.3.}
    \label{fig:3_1meanuvar}
\end{figure}

In table \ref{tab2}, the numerical displacement thickness $\delta^*_0$ refers to the displacement thickness of the laminar Blasius boundary layer imposed at the start of the numerical domain.
Notably, all data locations considered for the present analysis are sufficiently far away from the start of the computational domain ($x/{\delta^*_0}$ $\gtrsim$ 1500). 
Another important aspect to note in the LES-APG case is the steep increment in $\beta$, upstream of the nominally constant $\beta$ region, which is in contrast with the nominal ZPG condition maintained in case of Exp-APG dataset (figure \ref{fig:2_1tunnelschematic}b).
Hence, the statistics from LES-APG are inevitably influenced by the upstream-history effects.
This has been quantified by computing $\overline{\beta}$($x$), which indicates stronger history effects with increasing $x$-locations (\textit{i.e.} increasing $Re_{\tau}$).
As discussed, based on table \ref{tab0}, stronger history effects are known to increase small scale energy in the APG TBLs.
This has also been demonstrated in appendix 4 by comparing statistics from LES-APG, with the dataset of \citet{bobke2017history}, for similar $\beta$ and $Re_{\tau}$ but varying $\overline{\beta}$ (refer table \ref{tab2}). 
The trend of increasing small scale energy with $\overline{\beta}$, however, is not noted while comparing small scale statistics at increasing $Re_{\tau}$ ($\S$\ref{sec:4Re_effect}), suggesting history effects are weak and won't influence our conclusions associated with $Re_{\tau}$-effects, drawn from the LES-APG dataset.
Indeed, $\overline{\beta}$($x$ $=$ 6000${\delta}^*_{o}$) $\approx$ $\overline{\beta}$($x$ = 8500${\delta}^*_{o}$), reinforcing that history effects are weak towards the end of the computational domain.
Interested readers are referred to the respective articles \citep{eitel2014simulation,pozuelo2022} for further details on the numerical method, computational grid (${\Delta}{x^+}$,${\Delta}{z^+}$), etc. associated with the LES-ZPG and LES-APG datasets, respectively.

\section{Experimental data at high Reynolds-numbers}\label{sec:3exp_data}

This section presents the high-$Re_\tau$ ($\sim$ $6000$-$7000$) Exp-APG dataset, measured with the best available hotwire spatial-resolution ($l^+$ $\approx$ 10; table \ref{tab1}), and compares it with previously published data. 
Specifically, the mean velocity profiles, streamwise normal stresses and the energy spectrograms for $u$-fluctuations are compared for Exp-ZPG and Exp-APG datasets.
The idea is to demonstrate their consistency with trends noted previously in the literature \citep{monty2011,sanmiguel2020a,romero2022properties,aubertine2005} for varying $\beta$ in the low-to-moderate APG range (0 $\lesssim$ $\beta$ $\lesssim$ 2).

Figure \ref{fig:3_1meanuvar}(a) compares the viscous-scaled mean streamwise velocity ($U^+$) from Exp-APG and Exp-ZPG and compares it with published simulation datasets at low-$Re_{\tau}$, but in a similar $\beta$ range.
Consistent with the findings of \citet{nickels2004}, we note that the simulation data collapse for $z^+$ $<$ 10, which is made clearer by the enlarged plot shown in figure \ref{fig:3_1meanuvar}(c).
Similar collapse of the near-wall $U^+$ profiles for ZPG and APG TBL (with moderate $\beta$) was also noted by \citet{krogstad1995} and \citet{monty2011}, thereby forming a strong basis to estimate $U_{\tau}$ for the present experiments.
To this end, the mean velocity recorded from the near-wall hotwire measurements (0.2\:$mm$ $\lesssim$ $z$ $\lesssim$ 0.5\:$mm$), conducted for various $\beta$, are forced to fit over the published data as depicted in figure \ref{fig:3_1meanuvar}(c).
This fit is used to estimate $U_{\tau}$ for the Exp-ZPG and Exp-APG datasets presented in figure \ref{fig:3_1meanuvar}, which is recorded in table \ref{tab1}.
Notably, $U_{\tau}$ estimated for Exp-ZPG case matches well with the $U_{\tau}$ reported by \citet{hutchins2009}, reinforcing confidence in the methodology.

\begin{figure}[t]
    \centering
    \includegraphics[width=0.85\textwidth]{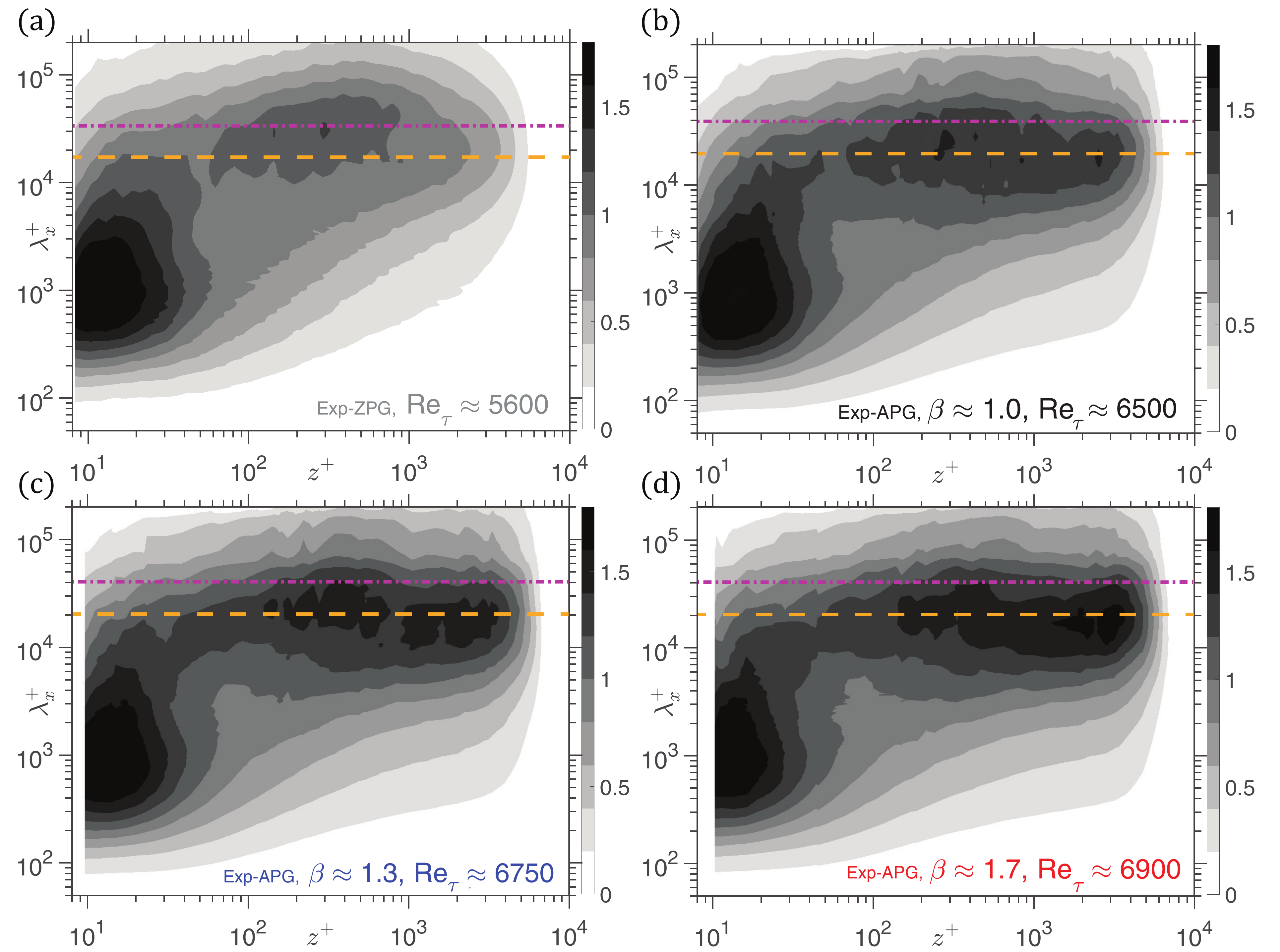}
    \caption{Viscous-scaled pre-multiplied energy spectra of the streamwise velocity fluctuations (${k_{x}}{{\phi}^{+}_{uu}}$), as a function of ${{\lambda}^{+}_{x}}$ and $z^+$, computed from the Exp-ZPG and Exp-APG datasets with $l^+$ $\approx$ 10 (table \ref{tab1}). 
    Dashed yellow and dash-dotted magenta lines indicate ${{\lambda}^{+}_{x}}$ = 3${\delta}^{+}_{99}$ and ${{\lambda}^{+}_{x}}$ = 6${\delta}^{+}_{99}$, respectively.}
\label{fig:3_2spectrograms}
\end{figure}

In the wake region (figure \ref{fig:3_1meanuvar}e), $U^+$ can be noted to increase in a way that suggests increasing size of the wake region with increasing $\beta$, which is expected \citep{skaare1994turbulent,monty2011}.
Next, we shift our focus to the intermediate $z^+$-range, ${\mathcal{O}}(10^2)$ $<$ $z^+$ $<$ ${\mathcal{O}}(10^3)$, where one could argue a logarithmic variation of $U^+$ vs $z^+$ (\textit{i.e.} the log-law).
This region, which is enlarged in figure \ref{fig:3_1meanuvar}(d) for clarity, suggests a systematic deviation of $U^+$ from the log-law as $\beta$ increases.
This observation is consistent with the findings of \citet{monty2011}, who also investigated $U^+$ versus $z^+$ trends for varying $\beta$ at matched $Re_{\tau}$.
Definitively determining whether this deviation is due to decreases in the value of the slope $\kappa$ \citep{knopp2021experimental} and/or a change in intercept `A', would require further careful investigation, while also accounting for the measurement uncertainty, and hence is out of scope (and focus) of the present study.
Moving to figure \ref{fig:3_1meanuvar}(b), the trends associated with ${\overline{u^2}}^{+}$ profiles obtained from Exp-ZPG and Exp-APG are also consistent with the literature \citep{monty2011,sanmiguel2020a}.
Energy in the outer region can be noted to increase significantly with $\beta$, with the ${\overline{u^2}}^{+}$ profile tending to plateau in the outer region for the largest $\beta$ $\approx$ 1.7.
The fact that no distinct outer peak can be noted in figure \ref{fig:3_1meanuvar}(b) is not surprising, and the plateauing of ${\overline{u^2}}^+$ profile in this $\beta$-range is consistent with the previous experimental studies at $Re_{\tau}$ $\gtrsim$ ${\mathcal{O}}$($10^3$) \citep{aubertine2005,monty2011,sanmiguel2020a,romero2022properties}, who noted a distinct outer peak only for $\beta$ $\gtrsim$ 2.
Streamwise normal stresses in the inner region, on the other hand, do not appear to change significantly in the $\beta$- and $Re_{\tau}$-range considered, which is also consistent with the studies of \citet{aubertine2005} and \citet{romero2022properties}. 

\begin{figure}[t!]
    \centering
    \includegraphics[width=1.0\textwidth]{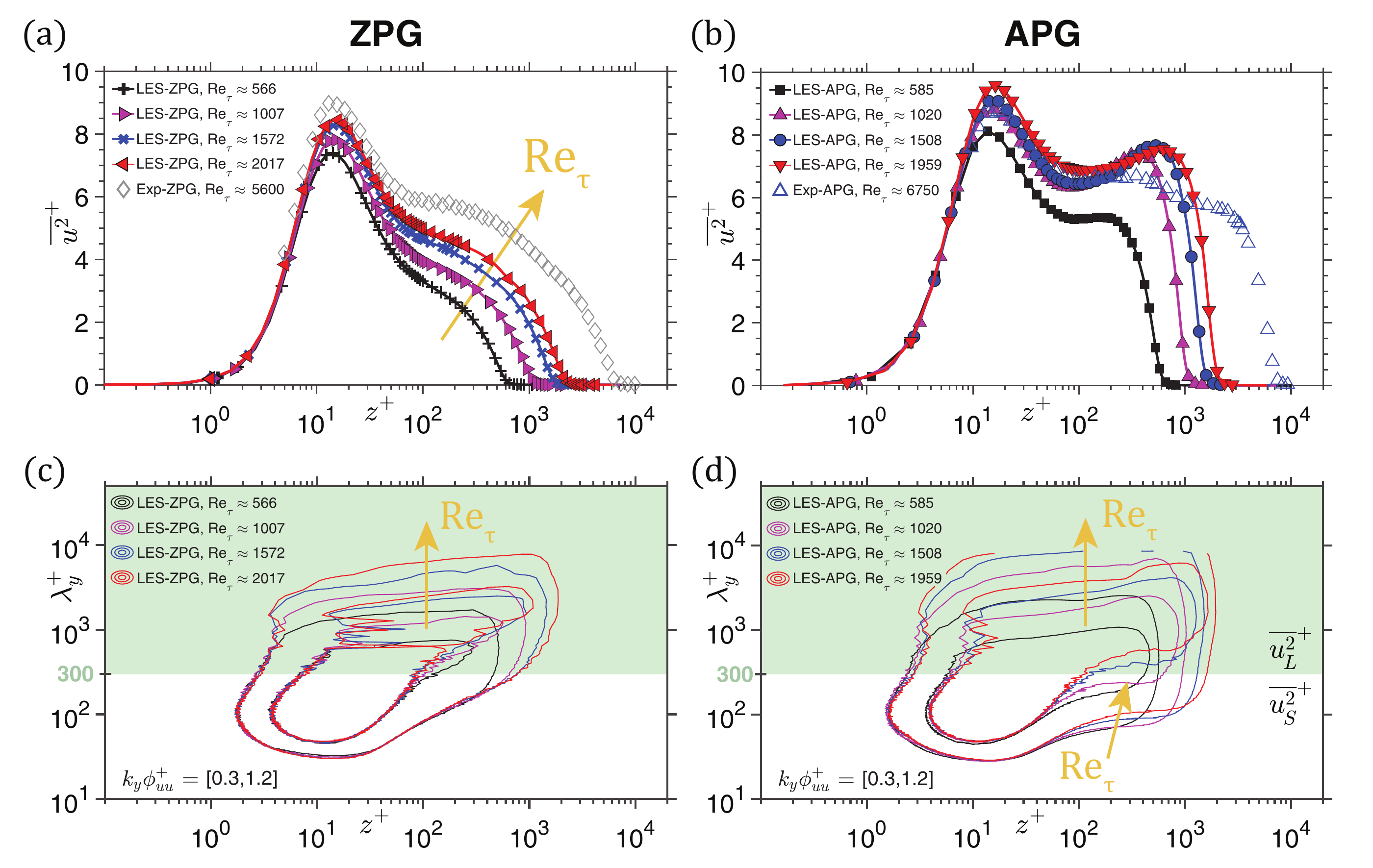}
    \caption{(a,b) Streamwise  normal stress, ${\overline{u^2}}^+$ and (c,d) constant energy contours of pre-multiplied $u$-spectra for (a,c) ZPG TBL and (b,d) APG TBL datasets at various $Re_{\tau}$. Estimates for LES-ZPG and LES-APG datasets correspond to their original (\textit{i.e.} well-resolved) grid resolution (${\Delta}{y^+}$; table \ref{tab2}) while those for Exp-ZPG and Exp-APG are from hotwires with $l^+$ $\approx$ 10 (table \ref{tab1}).
    Symbols and colour scheme for each dataset are as documented in $\S$\ref{sec:2setup}. 
    White and green background in (c,d) indicate small scale (${\overline{u^2_{S}}}^+$; ${\lambda}^{+}_{y}$ $\le$ 300) and large scale contributions (${\overline{u^2_{L}}}^+$; ${\lambda}^{+}_{y}$ $>$ 300), respectively.}
    \label{fig:4_1uvarspectra}
\end{figure}

In order to appreciate which turbulent scales contribute to the growth of ${\overline{u^2}}^{+}$ in the outer region, figure \ref{fig:3_2spectrograms} shows the $u$-spectrograms as a function of $z^+$ and ${\lambda}^{+}_{x}$ for the Exp-ZPG and Exp-APG datasets considered in figure \ref{fig:3_1meanuvar}.
Following \citet{harun2013} and \citet{sanmiguel2020a}, here we use Taylor's frozen turbulence hypothesis to convert time series to spatial length scales.
For this, we use the mean streamwise velocity at each $z^+$ as the mean convection velocity of all the turbulent scales coexisting at $z^+$. 
In general, variation in the scale-based energy distribution in the outer region, with $\beta$, is very similar to that noted previously by \citet{sanmiguel2020a} (refer their figure 3).
For the Exp-ZPG data (figure \ref{fig:3_2spectrograms}a), a characteristic outer peak can be noted at ${\lambda}^{+}_{x}$ $\sim$ 6${\delta}^+_{99}$ in the log-region \citep{hutchins2009}.
However, as $\beta$ increases to 0.9, scales coexisting beyond $z^+$ $>$ 10$^3$ get significantly more energetic than in Exp-ZPG, with energy concentrating at ${\lambda}^{+}_{x}$ $\sim$ 3${\delta}^{+}_{99}$ in this region.
This energy enhancement exists alongside the energy concentration noted at ${\lambda}^{+}_{x}$ $\sim$ 6${\delta}^+_{99}$, same as in Exp-ZPG.
As $\beta$ is increased further to 1.3 and 1.7, energy increases near both, ${\lambda}^{+}_{x}$ $\sim$ 3${\delta}^{+}_{99}$ and $\sim$ 6${\delta}^{+}_{99}$, consistent with \citet{harun2013} and \citet{sanmiguel2020a}.
Hence, based on figures \ref{fig:3_1meanuvar} and \ref{fig:3_2spectrograms}, we can conclude that the new Exp-APG dataset provides trends consistent with those in the literature, with larger values of $\beta$ increasing energy predominantly in the large scales.
This paves the way to use this dataset, with additional measurements at varying spatial-resolution $l^+$, for further analysis in $\S$\ref{sec:5spat_res}.

\section{Reynolds-number effect on the outer region of APG TBL}\label{sec:4Re_effect}

After establishing and discussing the quality of the Exp-APG dataset, we now compare and contrast it with the numerical datasets (LES-ZPG and LES-APG) to bring out the $Re_{\tau}$-trends associated with the streamwise normal stresses. 
For this, we begin by comparing ${\overline{u^2}}^+$ and the corresponding energy spectra in figure \ref{fig:4_1uvarspectra}.
The $Re_\tau$-trend observed for ${\overline{u^2}}^+$ of a canonical wall-bounded flow (figure \ref{fig:4_1uvarspectra}a) is already well-established in the literature \citep{hutchins2009,mklee2015}.
It involves a modest increase in energy of the inner-peak, located at $z^+ \approx 15$, and a significant growth of the outer region across 600 $\lesssim$ Re$_\tau$ $\lesssim$ 6000.
In the case of LES-APG, however, the variation of ${\overline{u^2}}^+$ in the outer region, for a similar $Re_{\tau}$ range, is indeterminate (figure \ref{fig:4_1uvarspectra}b).
A significant increase in the outer region, with emergence of an outer-peak, can be noted as $Re_{\tau}$ increases from 585$\rightarrow$1020.
However, this growth seems to stagnate with further increase in $Re_{\tau}$ across: 1000 $\lesssim$ $Re_{\tau}$ $\lesssim$ 2000.
For the purpose of comparison, figure \ref{fig:4_1uvarspectra}(b) also includes ${\overline{u^2}^+}$ profile obtained from the best-resolved hotwires ($l^+$ $\sim$ 10) for $\beta$ $\sim$ 1.3 and $Re_{\tau}$ $\approx$ 6750.
While it is not appropriate to make a quantitative comparison between the LES-APG and Exp-APG cases owing to their different upstream histories, the high-$Re_{\tau}$ Exp-APG data do indicate that  the indeterminate trend associated with ${\overline{u^2}}^+$ is not due to the relatively narrow $Re_{\tau}$ range of the LES-APG study.
The Exp-APG data also conforms to similar ${\overline{u^2}}^+$ levels in the outer region as LES-APG, which is in contrast with the trend noted from LES-ZPG and Exp-ZPG in the same $Re_{\tau}$ range.

\begin{figure}[t]
    \centering
    \includegraphics[width=0.95\textwidth]{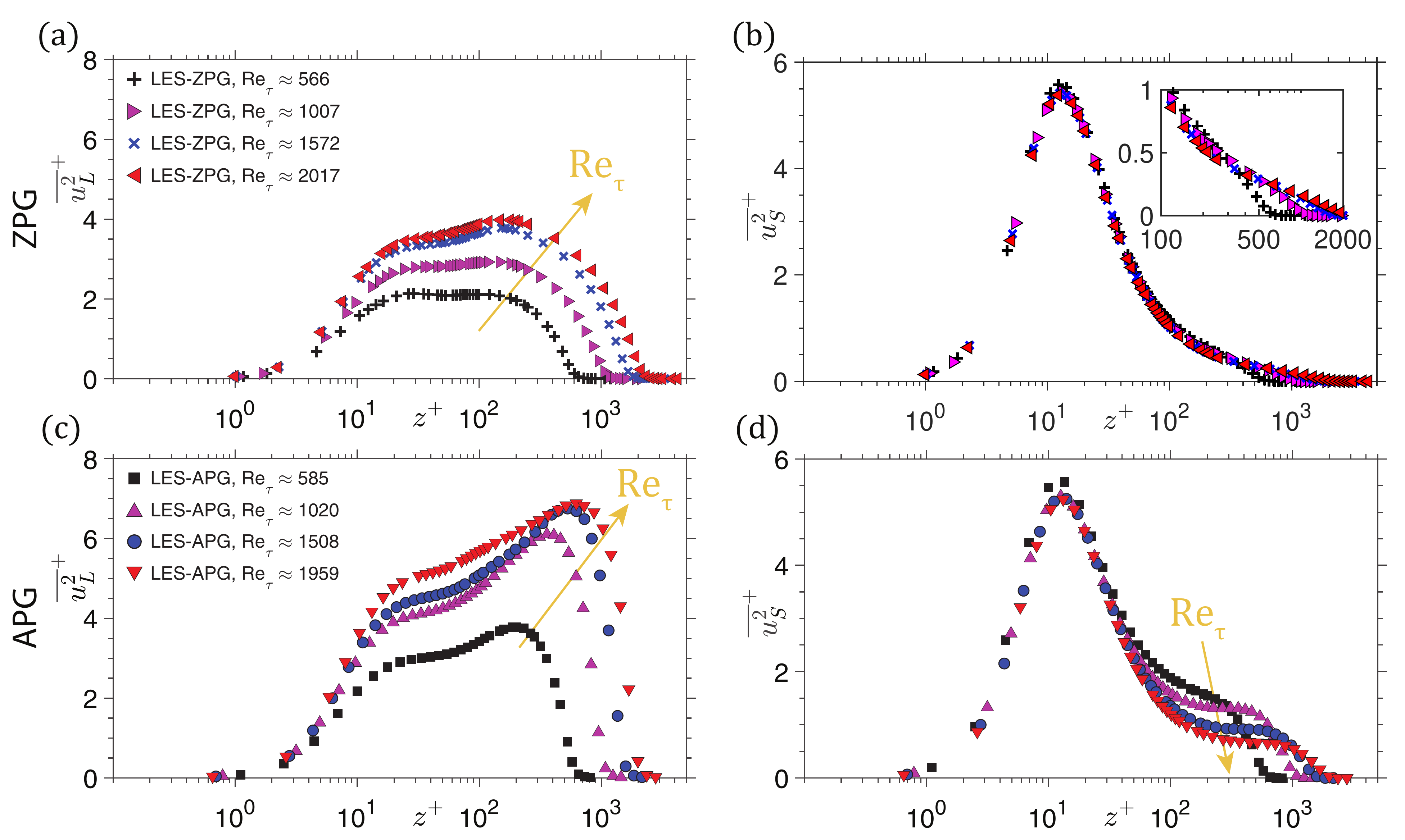}
    \caption{Streamwise normal stress decomposed into viscous-scaled (a,c) large scale (${\overline{u^2_{L}}}^+$) and (b,d) small scale contributions (${\overline{u^2_{S}}}^+$) for (a,b) LES-ZPG and (c,d) LES-APG datasets at various $Re_{\tau}$. Estimates for the simulation datasets correspond to their original (\textit{i.e.} well-resolved) grid resolution (${\Delta}{y^+}$; table \ref{tab2}).}
    \label{fig:4_2decomposition}
\end{figure}

To further examine the different trends for ${\overline{u^2}}^+$ in the outer region of APG and ZPG TBLs, we consider the constant-energy contours of $u$-energy spectra from LES-ZPG and LES-APG in figures \ref{fig:4_1uvarspectra}(c,d), respectively.
Here, we did not consider spectra from Exp-ZPG and Exp-APG datasets owing to unavailability of energy distributions as a function of ${\lambda}_{y}$.
In the case of LES-ZPG, the constant-energy contours exhibit a convincing collapse for the $u$-spectra in the small scale range across 600 $\lesssim$ $Re_{\tau}$ $\lesssim$ 2000.
Furthermore, energy in the large scales can be noted to increase with $Re_{\tau}$ \citep{hutchins2009}, as indicated by the outward shift of the contours towards higher ${\lambda}^{+}_{y}$.
As for the energy contours from LES-APG, the large scale energy can again be noted to increase with $Re_{\tau}$ \citep{tanarro2020}.
However, a contrasting trend is noted for the relatively small scales, where contours are noted to shift inwards, indicating a decrease in energy with $Re_{\tau}$.

To clearly bring out the $Re_{\tau}$-trends of the small and large scale energy from LES-ZPG and LES-APG, we consider decomposing the total streamwise variance (${\overline{u^2}}^+$) into its large (${\overline{u^2_{L}}}^+$) and small  (${\overline{u^2_{S}}}^+$) energy contributions.
To this end, we use a sharp spectral cut-off at ${\lambda}^{+}_{y,c}$ $=$ 300 to decompose ${\overline{u^2}}^+$ = ${\overline{u^2_{S}}}^+$ + ${\overline{u^2_{L}}}^+$.
Here, a constant viscous-scaled cut-off has been chosen based on a careful consideration of published datasets of canonical wall-bounded flows \citep{sillero2014,eitel2014simulation,mklee2015,mklee2019}, to segregate a fixed range of small (viscous) scales.
The ${\overline{u^2_{S}}}^+$ distributions obtained from these datasets exhibited a quasi Re$_{\tau}$-invariance \citep{hutchins2009} for ${\lambda}^{+}_{y_c}$ $\lesssim$ 300, a fact that was used as a basis to select ${\lambda}^+_{y_c}$ = $300$ \citep{mklee2019}.
This has been demonstrated for the DNS dataset of turbulent channel flow \citep{mklee2015} in appendix 3, as well as for the LES-ZPG dataset in figure \ref{fig:4_2decomposition}(b).
It is, however, important to note that the present conclusions do not depend on the exact choice of ${\lambda}^+_{y_c}$, which is used to simply reveal the contrasting trends in ${\overline{u^2_{S}}}^+$ for LES-ZPG and LES-APG.
Figures \ref{fig:4_2decomposition} shows ${\overline{u^2_{L}}}^+$ and ${\overline{u^2_{S}}}^+$ obtained using ${\lambda}^+_{y_c}$ $=$ 300 for LES-ZPG and LES-APG datasets.
These plots now clearly show the trends discussed based on the spectra in figure \ref{fig:4_1uvarspectra}(c,d): ${\overline{u^2_{L}}}^+$ can be noted to increase across the entire TBL for both LES-APG and LES-ZPG, and is representative of the increase in energy of the inertial scales with increasing $Re_{\tau}$ (\textit{i.e.} Townsend's attached eddies and superstructures; \citep{smits2011}).
While ${\overline{u^2_{S}}}^+$ is quasi $Re_{\tau}$-invariant for LES-ZPG across entire TBL, and in the near-wall region for LES-APG, which would be expected for the fine dissipative and viscous scales associated with this small scale range \citep{hutchins2009}.
The inset in figure \ref{fig:4_2decomposition}(b) confirms that the $Re_{\tau}$-invariance of ${\overline{u^2_{S}}}^+$ extends up to as high as $z^+$ $\lesssim$ 0.8$Re_{\tau}$, beyond which ${\overline{u^2_{S}}}^{+}$ gradually decreases to zero owing to the increasing influence of the turbulent/non-turbulent interface of the TBL.
Interestingly, however, ${\overline{u^2_{S}}}^+$ for the LES-APG can be noted to decrease with $Re_{\tau}$ in the outer region, which is consistent with our hypothesis discussed in $\S$\ref{sec:1intro}.
The contrasting trends exhibited by ${\overline{u^2_{L}}}^+$ and ${\overline{u^2_{S}}}^+$ in the outer region for LES-APG explains the indeterminate variation of ${\overline{u^2}}^+$, with $Re_{\tau}$, in figure \ref{fig:4_1uvarspectra}(b).

\begin{figure}[t]
    \centering
    \includegraphics[width=1.0\textwidth]{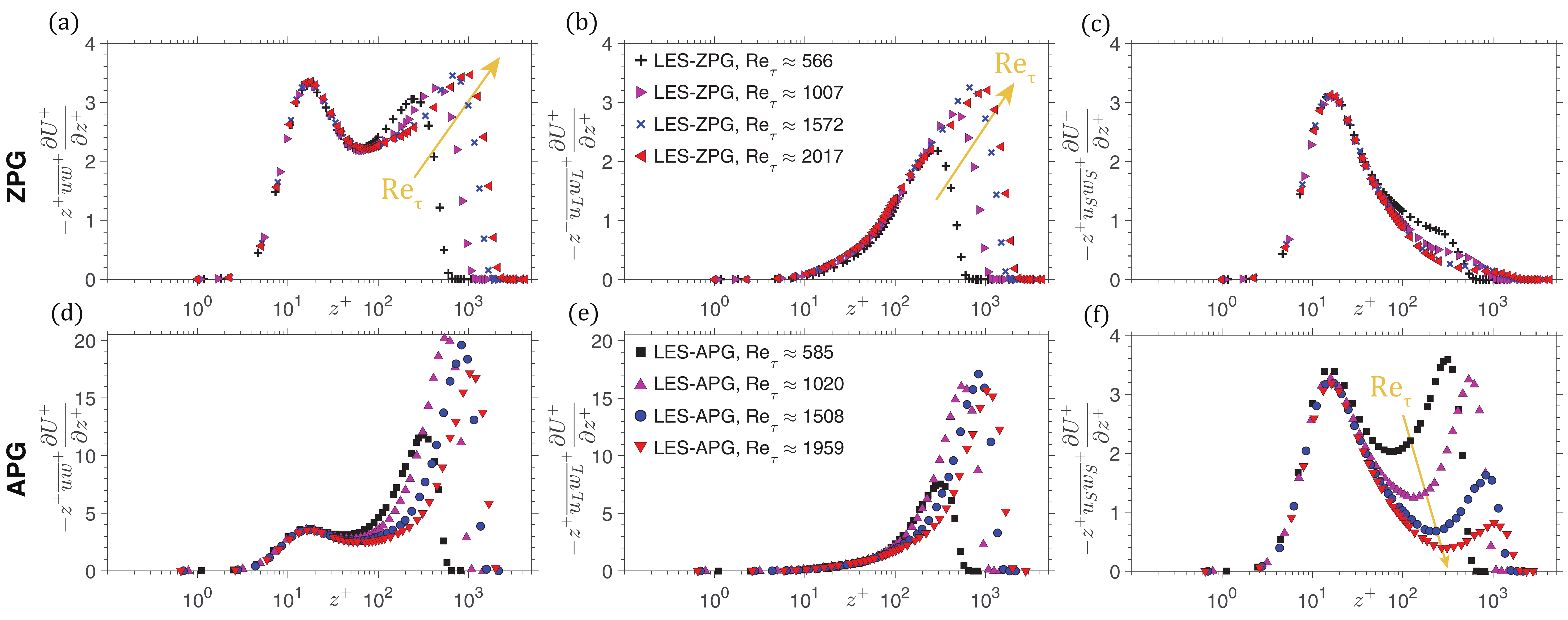}
    \caption{(a,d) Dominant contribution to production of turbulent-kinetic-energy, $-{{\overline{uw}}^{+}}{{\partial}U^+ / {\partial}z^+}$, decomposed into viscous-scaled (b,e) large scale ($-{{\overline{{u_{L}}{w_{L}}}}^+}{{\partial}U^+ / {\partial}z^+}$) and (c,f) small scale contributions ($-{{\overline{{u_{S}}{w_{S}}}}^+}{{\partial}U^+ / {\partial}z^+}$) for (a-c) LES-ZPG and (d-f) LES-APG datasets at various $Re_{\tau}$. Estimates for the simulation datasets correspond to their original grid resolution (${\Delta}{y^+}$; table \ref{tab2}).
    Note that the profiles are pre-multiplied with $z^+$. Also, the range of the vertical axis for (a,b,c,f) is different than for (d,e).}
    \label{fig:4_3production}
\end{figure}

\begin{figure}[t!] 
    \centering
    \includegraphics[width=0.94\textwidth]{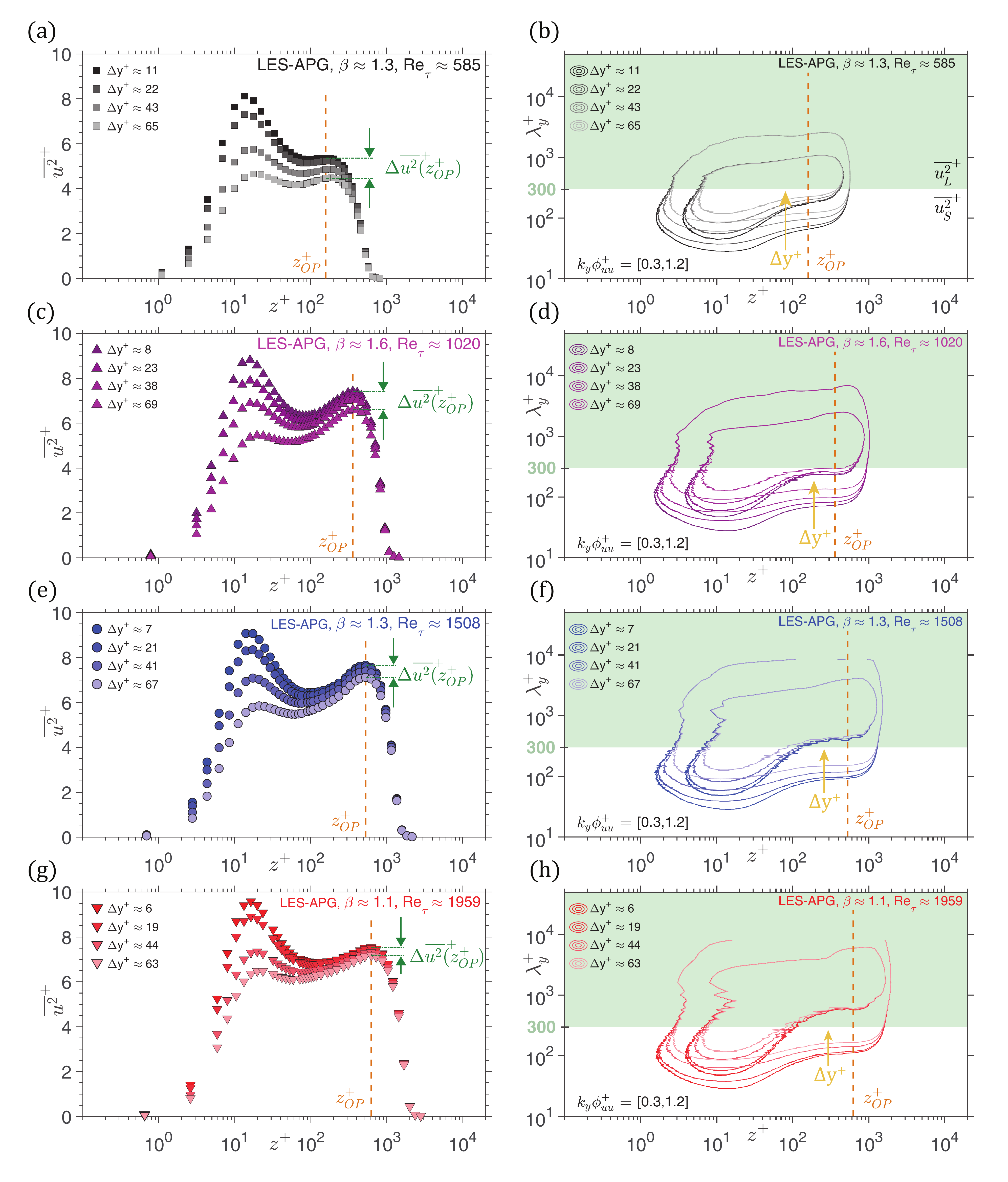}
    \caption{(a,c,e,f) Streamwise normal stress, ${\overline{u^2}}^+$ and (b,d,f,h) constant-energy contours of pre-multiplied $u$-spectra estimated for LES-APG datsets with various spatial-resolutions (${\Delta}{y^+}$) at $Re_{\tau}$ $\approx$ (a,b) 585, (c,d) 1020, (e,f) 1508 and (g,h) 1959.
    Symbols and colour scheme for each dataset are as documented in table \ref{tab2}. 
    Dashed orange lines indicate $z^+$ associated with the outer-peak (\textit{i.e.} $z^{+}_{OP}$). 
    ${\Delta}{\overline{u^2}^+}(z^+_{OP})$ in (a,c,e,g) corresponds to the attenuation in ${{\overline{u^2}}^+}$ at $z^+_{OP}$ as ${\Delta}{y^+}$ increases from the well-resolved to worst-resolved scenario.
    White and green background in (b,d,f,h) indicate small scale (${\overline{u^2_{S}}}^+$; ${\lambda}^{+}_{y}$ $\le$ 300) and large scale contributions (${\overline{u^2_{L}}}^+$; ${\lambda}^{+}_{y}$ $>$ 300), respectively.}
    \label{fig:5_1SR_numerical}
\end{figure}

To understand the origins of the decrease in small scale energy with $Re_{\tau}$, we consider the turbulent-kinetic-energy budget.
Specifically, we investigate the production of the total turbulent-kinetic-energy, $k^+$ = (${\overline{u^2}^+}$ + ${\overline{v^2}^+}$ + ${\overline{w^2}^+}$)/2, given by $P^+_{k}$ $=$ -${\overline{{u_i}{u_j}}}^+$(${{\partial}U^+_i}/{{\partial}x^+_j}$) \citep{bobke2017history,pozuelo2022}, where $i$ and $j$ represent the conventional tensor notations.
In the case of APG TBLs, $P^+_{k}$ $=$ -${\overline{{u^2}}}^+$(${{\partial}U^+}/{{\partial}x^+}$) -${\overline{{w^2}}}^+$(${{\partial}W^+}/{{\partial}z^+}$) -${\overline{{u}{w}}}^+$(${{\partial}U^+}/{{\partial}z^+}$ + ${{\partial}W^+}/{{\partial}x^+}$), out of which -${\overline{{u}{w}}}^+$(${{\partial}U^+}/{{\partial}z^+}$) has been found to be much larger than the other three terms \citep{pozuelo2022,gungor2022}, at least in case of moderate $\beta$. 
Also note that -${\overline{{u}{w}}}^+$(${{\partial}U^+}/{{\partial}z^+}$) is also the only non-zero term when $P^{+}_{k}$ is expanded for the case of canonical wall flows \citep{marusic2010high,mklee2019}, and is essentially the product of the Reynolds shear stress (${\overline{{u}{w}}}^+$) and mean velocity gradient (${{\partial}U^+}/{{\partial}z^+}$) at any $z^+$.
Hence, we limit our attention to this term in the present study and analyze its variation with $Re_{\tau}$ using LES-ZPG and LES-APG datasets.
We also decompose -${\overline{{u}{w}}}^+$(${{\partial}U^+}/{{\partial}z^+}$) into dominant contributions from the small (-${\overline{{u_{s}}{w_{s}}}}^+$(${{\partial}U^+}/{{\partial}z^+}$)) and large scales (-${\overline{{u_{L}}{w_{L}}}}^+$(${{\partial}U^+}/{{\partial}z^+}$)), by using the same spectral cut-off (${\lambda}^{+}_{y,c}$ = 300) to segregate $w$ $=$ $w_S$ + $w_{L}$.

Figure \ref{fig:4_3production} shows these three terms in a pre-multiplied form following \citet{marusic2010high}, so that equal areas under the graph represent equal integral contributions.
In the case of LES-ZPG, the $Re_{\tau}$-trends are again consistent with the literature \citep{marusic2010high,mklee2019}, where increasing $Re_{\tau}$ increases the production of turbulent-kinetic-energy in the outer region.
This can be associated with increasing production in the large scales, while variation in the small scale characteristics is negligible.
In comparison to LES-ZPG, the production in the outer region of LES-APG is enhanced significantly, which is also well-known in the literature \citep{skaare1994turbulent,bobke2017history,pozuelo2022,gungor2022}.
However, it is evident from figure \ref{fig:4_3production}(d) that the outer peak in -${\overline{{u}{w}}}^+$(${{\partial}U^+}/{{\partial}z^+}$) appears to be reducing with $Re_{\tau}$, opposite to the trend for LES-ZPG.
While this observation was discussed in \citet{pozuelo2022}, here we can associate this decrease partly to the decrease in production of turbulence in the small scales with $Re_{\tau}$, represented by -${\overline{{u_{s}}{w_{s}}}}^+$(${{\partial}U^+}/{{\partial}z^+}$) in figure \ref{fig:4_3production}(f).
This reduction in the turbulence production, thus, gives a physical explanation for the drop in ${\overline{u^2_{S}}}^+$ with increasing $Re_{\tau}$, in figure \ref{fig:4_2decomposition}(d).
The fact that this decrease in ${\overline{u^2_{S}}}^+$  and -${\overline{{u_{s}}{w_{s}}}}^+$(${{\partial}U^+}/{{\partial}z^+}$) in the outer region is purely a $Re_{\tau}$-effect, and not an artefact of increasing $\overline{\beta}$ (table \ref{tab2}), has been confirmed in appendix 4. 
Similarly, we can argue that this decrease in small-scale contributions is not an artefact of mild changes in $\beta$; the argument is based on the fact that the trend is consistent despite the contrasting variations in $\beta$ (table \ref{tab2}), which slightly increases for 585 $\lesssim$ $Re_{\tau}$ $\lesssim$ 1020, is equal for ${Re_{\tau}}{\sim}$585 and 1508, after which it reduces for ${Re_{\tau}}{\sim}$1959.
It is worth noting in the case of figure \ref{fig:4_3production} that the spectral decomposition of -${\overline{{u}{w}}}^+$(${{\partial}U^+}/{{\partial}z^+}$) also leads to two other terms, -${\overline{{u_{S}}{w_{L}}}}^+$(${{\partial}U^+}/{{\partial}z^+}$) and -${\overline{{u_{L}}{w_{S}}}}^+$(${{\partial}U^+}/{{\partial}z^+}$), which however are much smaller than the two main terms plotted in the figure (and are hence, neglected).

\section{Spatial-resolution effect for varying Reynolds-number}\label{sec:5spat_res}

\begin{figure}[t]
    \centering
    \includegraphics[width=0.6\textwidth]{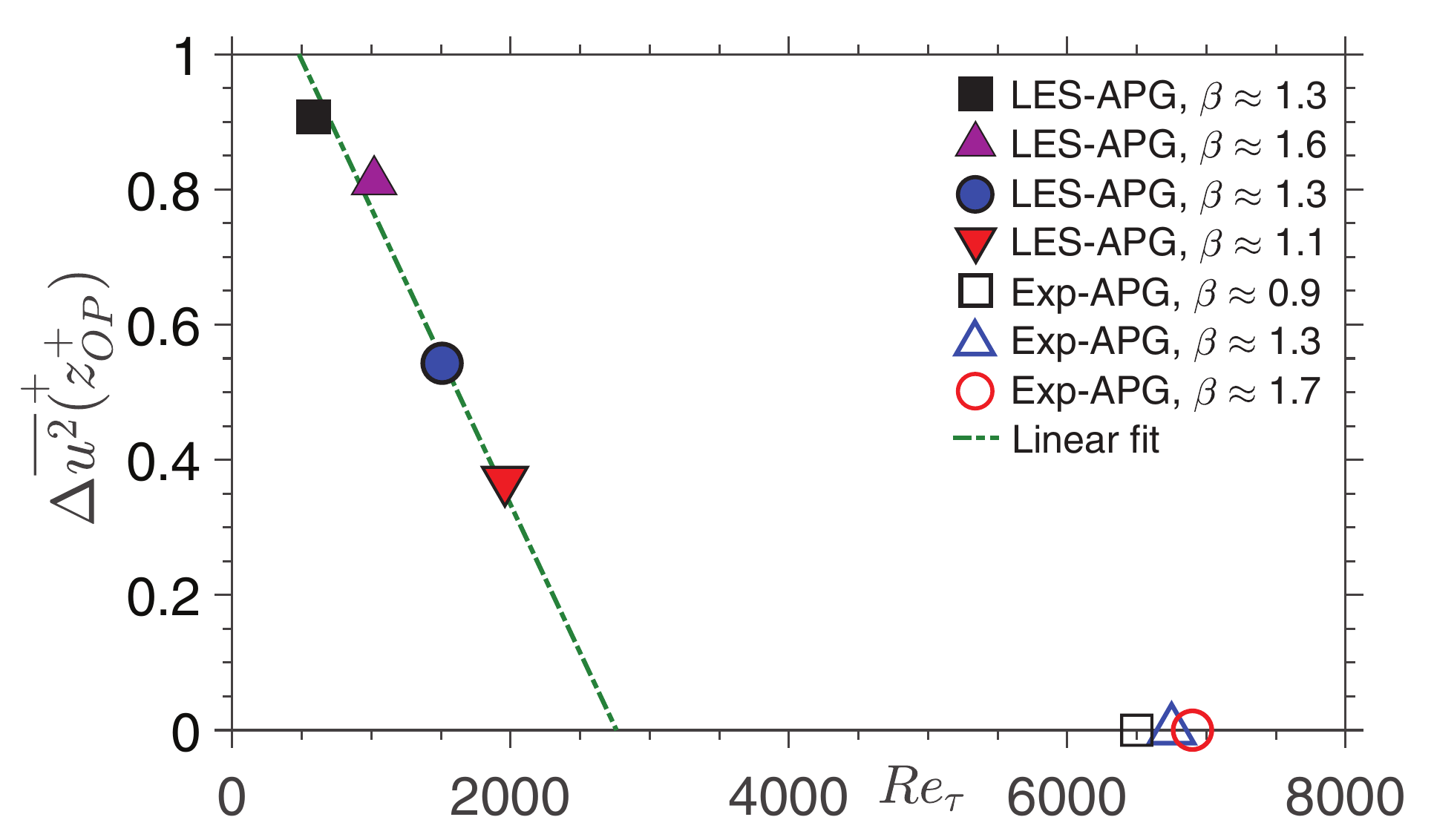}
    \caption{Variation in attenuation of ${{\overline{u^2}}^+}$ at $z^+_{OP}$ (i.e ${\Delta}{\overline{u^2}^+}(z^+_{OP})$) with $Re_{\tau}$. Here, ${\Delta}{\overline{u^2}^+}(z^+_{OP})$ is the difference between ${{\overline{u^2}}^+}$($z^+_{OP}$) as ${\Delta}{y^+}$ increases from the well-resolved to worst-resolved scenario (tables \ref{tab1},\ref{tab2}). Note that $z^+_{OP}$ is assumed at 0.3${\delta}^{+}_{99}$ for the Exp-APG data following \citet{sanmiguel2020a}.
    Dashed green line indicates linear fit based purely on the LES-APG data.}
    \label{fig:5_2SRerror}
\end{figure}

The analysis presented in $\S$\ref{sec:4Re_effect} concludes that the turbulent-kinetic-energy of the small scales, in the outer region of an APG TBL, reduces with increasing $Re_{\tau}$.
This will have ramifications for the measurement errors arising due to spatial-resolution issues, which we know are associated with attenuation of the small scale energy contributions \citep{hutchins2009}. 
A decrease in small scale energy with increasing $Re_{\tau}$, noted here for an APG TBL, should reduce the attenuation in ${\overline{u^2}}^+$ profiles estimated for various spatial-resolutions (with increasing $Re_{\tau}$).
This is tested in figure \ref{fig:5_1SR_numerical}, which compares LES-APG ${\overline{u^2}}^+$ profiles and the corresponding $u$-spectra, estimated for various spatial-resolutions (${\Delta}{y^+}$), as documented in table \ref{tab2}.
The spatial-resolution is systematically varied for each $Re_{\tau}$ case in the range 6 $\lesssim$ ${\Delta}{y^+}$ $\lesssim$ 69.
Since our focus is on the outer region, the outer peak ($z^{+}_{OP}$) associated with each LES-APG case is identified from well-resolved data and marked in figure \ref{fig:5_1SR_numerical} as a dashed line.
It is evident from figure \ref{fig:5_1SR_numerical} that an increase in ${\Delta}{y^+}$ attenuates the estimated normal stress, ${\overline{u^2}}^+$.
However, the degree of attenuation varies across the TBL thickness as well as for each $Re_{\tau}$.
To quantify this variation in attenuation for the outer region, we compute the difference in ${\overline{u^2}}^+$ for the well-resolved and worst-resolved case at $z^{+}_{OP}$ for each $Re_{\tau}$.
This is defined as ${\Delta}{{\overline{{u^2}}}^{+}}(z^{+}_{OP})$ for each LES-APG case in figure \ref{fig:5_1SR_numerical} and plotted in figure \ref{fig:5_2SRerror} as a function of $Re_{\tau}$.
It is evident that the energy attenuation (${\Delta}{{\overline{{u^2}}}^{+}}(z^{+}_{OP})$) owing to increasing ${\Delta}{y^+}$ reduces with $Re_{\tau}$.
The constant-energy contours of the $u$-spectra, plotted in figures \ref{fig:5_1SR_numerical}(b,d,f,h) for different $Re_{\tau}$, confirm that this energy attenuation occurs solely in the small scale range (${\lambda}^{+}_{y}$ $\lesssim$ 300), as expected.
Hence, the spectra not only support the usage of ${\lambda}^{+}_{y,c}$ = 300 for decomposing ${\overline{u^2}}^+$ into ${\overline{u^2_{L}}}^+$ and ${\overline{u^2_{S}}}^+$, but they also indicate that the decreasing trend of ${\Delta}{{\overline{{u^2}}}^{+}}(z^{+}_{OP})$ with $Re_{\tau}$ is solely associated with reduced attenuation of the small scale energy with $Re_{\tau}$.
This, in turn, is associated with the reduction in small scale energy in the outer region with $Re_{\tau}$ (figure \ref{fig:4_1uvarspectra}d).

\begin{figure}[t]
    \centering
    \includegraphics[width=1.05\textwidth]{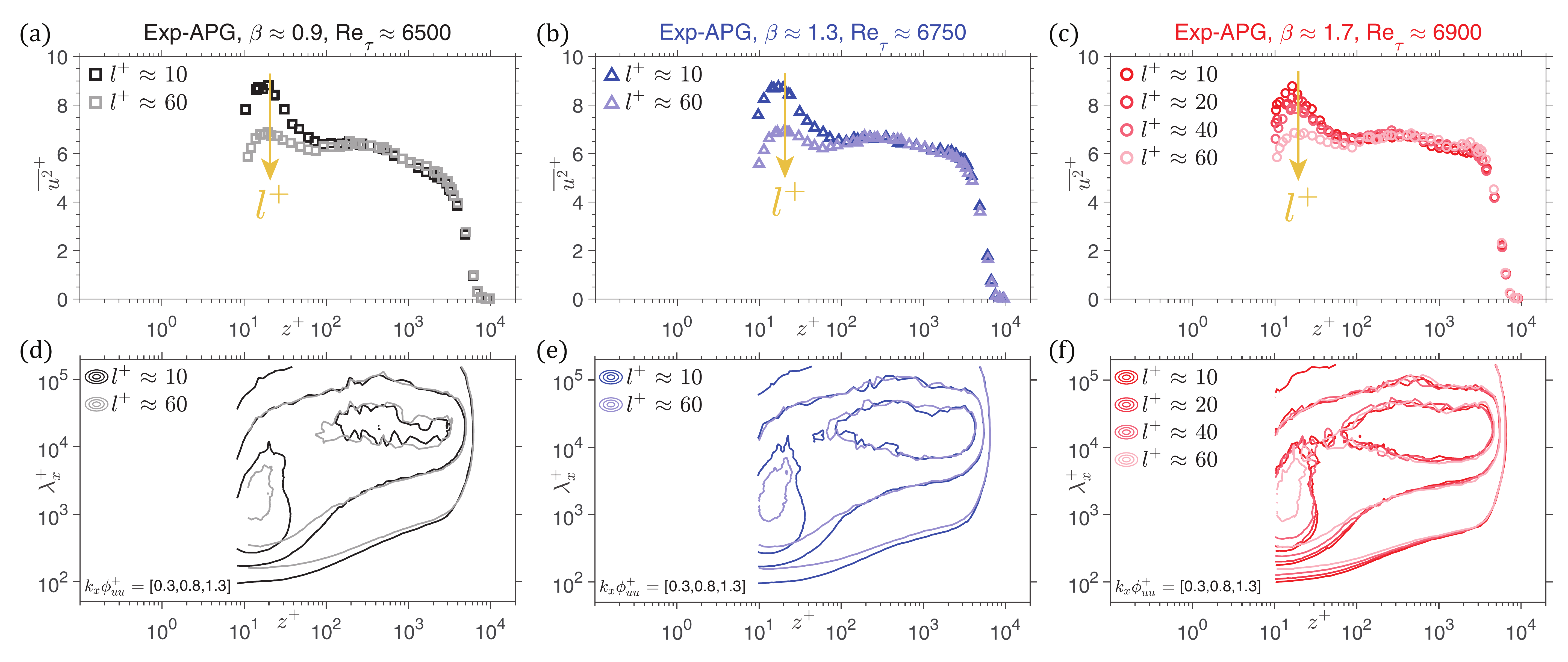}
    \caption{(a-c) Streamwise turbulent-kinetic-energy, ${\overline{u^2}}^+$ and (d-f) constant-energy contours of pre-multiplied $u$-spectra estimated for Exp-APG datasets with various spatial-resolutions ($l^+$) at $\beta$ $\approx$ (a,d) 0.9, (b,e) 1.3 and (c,f) 1.7.
    Symbols and colour scheme for each dataset are as documented in table \ref{tab1}.}
    \label{fig:5_3SR_experimental}
\end{figure}

The trend associated with reduction of ${\Delta}{{\overline{{u^2}}}^{+}}(z^{+}_{OP})$ with increasing $Re_{\tau}$, in figure \ref{fig:5_2SRerror}, can be used to estimate approximate $Re_{\tau}$ range when ${\Delta}{{\overline{{u^2}}}^{+}}(z^{+}_{OP})$ $\rightarrow$ 0.
The simplest way is to fit a straight line, which indicates ${\Delta}{{\overline{{u^2}}}^{+}}(z^{+}_{OP})$ $\approx$ 0 for $Re_{\tau}$ $\gtrsim$ 3000.
It should be noted, however, that this is a very crude estimate based on a very limited parameter range ($\beta$, $\overline{\beta}$), and is also specific to the spatial-resolution associated with well-resolved (${\Delta}{y^+}$ $\sim$ 6) and worst-resolved (${\Delta}{y^+}$ $\sim$ 69) cases considered here.
This prediction can be tested in the present study by using the high $Re_{\tau}$ Exp-APG dataset, acquired for a similar range of spatial-resolution (10 $\lesssim$ $l^+$ $\lesssim$ 60) and moderate $\beta$ range as the LES-APG.
Figure \ref{fig:5_3SR_experimental} depicts ${\overline{u^2}}^+$ and constant-energy contours of the associated $u$-energy spectra from the Exp-APG dataset for various $\beta$ and spatial-resolutions.
Remarkably, ${\overline{u^2}}^+$ can be noted to attenuate with increasing $l^+$ only in the inner region ($z^+$ $<$ 100), with negligible changes in the outer region.
Although there is no clear outer peak evident in ${\overline{u^2}}^+$ for Exp-APG data, for the purpose of comparison, we estimate ${\Delta}{{\overline{{u^2}}}^{+}}(z^{+}_{OP})$ for this data by using $z^{+}_{OP}$ $\sim$ 0.3${\delta}^{+}_{99}$ following \citet{sanmiguel2020a} and plot it in figure \ref{fig:5_2SRerror}.
Indeed, ${\Delta}{{\overline{{u^2}}}^{+}}(z^{+}_{OP})$ $\sim$ 0 at $Re_{\tau}$ $\gtrsim$ 6500 associated with Exp-APG, consistent with the crude prediction from figure \ref{fig:5_2SRerror}.

The $Re_{\tau}$-trend in figure \ref{fig:5_2SRerror} can be explained based on the fact that the small scale energy in the outer region reduces significantly as $Re_{\tau}$ $\gtrsim$ 6500.
This is confirmed from the $u$-spectra contours associated with Exp-APG datasets in figures \ref{fig:5_3SR_experimental}(d-f), which indicate very low energy at small ${\lambda}^{+}_{x}$ in the outer region.
Consequently, increasing ${\Delta}{y^+}$ only attenuates the small scale energy in the inner region and negligibly influences the contours in the outer region.
Apart from reaffirming the $Re_{\tau}$-trend of the small scale energy, the analysis presented in figures \ref{fig:5_1SR_numerical}-\ref{fig:5_3SR_experimental} also suggests negligible influence of spatial-resolution effects ($l^+$ $\lesssim$ 60) in the outer region of high $Re_{\tau}$ APG TBLs ($\gtrsim$ 3000), at least for moderate $\beta$ ($\lesssim$ 2).
This is reassuring for past as well as future experimental studies investigating this parameter space, especially from the perspective of proposal and validation of new scaling arguments for the outer region.
Present conclusion, however, adds additional challenges for correction schemes for APG TBLs (i.e. dependence on $Re_{\tau}$, $\beta$ and $\overline{\beta}$), which in the case of ZPG TBLs have relied on the negligible small scale energy variation with $Re_{\tau}$ \citep{chin2011,jhlee2016,deshpande2020}.
It is worth noting here that the present Exp-APG dataset was acquired in a TBL with a well-controlled upstream-history.
Hence, it is possible that the negligible spatial-resolution effects noted here at $Re_{\tau}$ $\sim$ 7000 may be associated with the given upstream-history.
However, the $Re_{\tau}$-trend demonstrated based on the LES-APG dataset confirms that the spatial-attenuation in the outer region does reduce with $Re_{\tau}$, and this is related to the reduction in small scale energy with $Re_{\tau}$.
Another aspect worth discussing here is the noticeable scatter in the ${\overline{u^2}}^+$ profiles observed in figure \ref{fig:5_3SR_experimental}(c), for Exp-APG data at $\beta$ $\approx$ 1.7.
Based on close inspection of the repeated experiments at various $l^+$, we believe that this is owing to measurement uncertainty, which may reduce by increasing data-acquisition time, \textit{i.e.} $T{U_{\infty}}/{{\delta}_{99}}$ $>$ 20000.
This, however, was not tested here due to the already very long acquisition time ($T$ $=$ 720\:$s$) used for each $z^+$ location in Exp-APG, beyond which hotwire drift will likely be significant.
This could be tested in future studies conducted at higher free-stream velocites.

\section{Concluding remarks}\label{sec:6conclusion}

The present study investigates $Re_{\tau}$ effects in the outer region of an APG TBL by analyzing datasets across a decade of $Re_{\tau}$ (600 $\lesssim$ $Re_{\tau}$ $\lesssim$ 7000).
The data is a combination of the well-resolved LES dataset of \citet{pozuelo2022}, corresponding to the low-$Re_{\tau}$ range, and new high-$Re_{\tau}$ experimental data from the Melbourne wind tunnel, both of which correspond to moderate-$\beta$ range ($\lesssim$ 2).
A unique aspect of the LES dataset is the considerable $Re_{\tau}$ range, 1000 $\lesssim$ $Re_{\tau}$ $\lesssim$ 2000, along which the APG TBL is allowed to develop at a nominally constant $\beta$ $\sim$ 1.4. 
This enables investigation of the $Re_{\tau}$ effects with negligible influence of $\beta$ and a very weak change in the upstream-history effects with $x$ ($\overline{\beta}$($x$)).
In the case of the experimental data, on the other hand, the streamwise pressure gradient is set up based on the methodology of \citet{clauser1954}, wherein the APG TBL grows starting from a canonical TBL at $Re_{\tau}$ $\sim$ 4000, \textit{i.e.} with no unique upstream-history effects.
Analysis on both this data is complemented by using published datasets of ZPG TBLs, in the same $Re_{\tau}$ range, to identify the unique trends associated with APG TBLs.

Scale-based decomposition of the streamwise normal stresses reveals that the viscous-scaled small scale energy (${\overline{u^2_{S}}}^+$) in the outer region of an APG TBL, decreases with $Re_{\tau}$, which is in contrast with quasi-$Re_{\tau}$ invariance depicted by these scales in ZPG TBLs. 
This conclusion provides a basis of support for the hypothesis of \citet{tanarro2020} that increasing Reynolds numbers lead to a decreasing influence of the APG effects.
On the other hand, the large scale energy (${\overline{u^2_{L}}}^+$) corresponding to attached eddies and superstructures is found to increase with $Re_{\tau}$ across the entire TBL, for both the ZPG and APG case.
The contrasting $Re_{\tau}$ trends for ${\overline{u^2_{S}}}^+$ and ${\overline{u^2_{L}}}^+$, in the case of APG TBLs, is deemed responsible for the indeterminate trend of ${\overline{u^2}^+}$ profiles with $Re_{\tau}$ (figure \ref{fig:4_1uvarspectra}b).
To understand the origins of this phenomenon, the dominant term associated with production of turbulent kinetic energy, ${\overline{{u}{w}}}^+$(${{\partial}U^+}/{{\partial}z^+}$), is investigated across 600 $\lesssim$ $Re_{\tau}$ $\lesssim$ 2000.
The analysis reveals that production of turbulence in the small scales decreases with $Re_{\tau}$ in the outer region of an APG TBL, thereby explaining the decrease in ${\overline{{u^2_{S}}}}^+$.
This decrease in small scale production also (partly) explains the drop in the outer peak of the net production of turbulent kinetic energy, noted previously by \citet{pozuelo2022}.

The $Re_{\tau}$ effects on the small-scale energy are also confirmed by investigating the attenuation of ${\overline{{u^2}}}^+$ by decreasing spanwise resolution (i.e. increasing ${\Delta}{y^+}$ or $l^+$), as a function of $Re_{\tau}$.
The degree of attenuation of ${\overline{{u^2}}}^+$ quantified at the outer peak location $z^+_{OP}$, albeit significant at low-$Re_{\tau}$ ($\sim$ 1000), was found to decrease with $Re_{\tau}$ in a quasi-linear manner.
The trend suggested negligible energy attenuation in the outer region for $Re_{\tau}$ $\gtrsim$ 3000, which was tested by using hotwire data acquired for the nominally same spatial-resolution range.
Experimental data at $Re_{\tau}$ $\gtrsim$ 6500 confirmed negligible attenuation in ${\overline{{u^2}}}^+$ in the outer region of a moderate APG TBL, reaffirming the $Re_{\tau}$ effects on the small scale energy concluded based on energy decomposition analysis.
It would be interesting to assess whether a similar reduction in small scale energy with $Re_{\tau}$ is also noted for much higher $\beta$ (where the flow characteristics are closer to a wall-confined wake \citep{maciel2018outer,gungor2022,kitsios2016direct}), and if these contributions become negligible at a much larger $Re_{\tau}$ than present predictions.
This study, hence, informs future studies requiring high-$Re_{\tau}$ APG TBLs with negligible upstream-history effects.
Potential projects include rigorously investigating changes in the log-law (\textit{i.e.} $\kappa$ and $A$,\citep{knopp2021experimental}, which likely would require higher-fidelity $U_\tau$ measurements) and testing scaling arguments for the outer peak \citep{pozuelo2022,pozuelo2023,wei2023} based on high-$Re_{\tau}$ data ($\gtrsim$ ${\mathcal{O}}$(10$^4$)).
Present work confirms that high-$Re_{\tau}$ experiments in the outer region would have negligible errors due to spanwise spatial-resolution (for moderate $\beta$ $\lesssim$ 2 and modest $l^+$ $\lesssim$ 60), which is good news for these future experiments.
Decrease in small scale energy in the outer region with increasing $Re_{\tau}$, however, suggests that any new scaling arguments for APG TBLs should be rigorously tested across a broad range of $Re_{\tau}$.

\section*{Appendix 1: Confirmation of `canonical' upstream condition for Exp-APG}\label{app:1upstreamcheck}

\begin{figure}[t]
    \centering
    \includegraphics[width=1.0\textwidth]{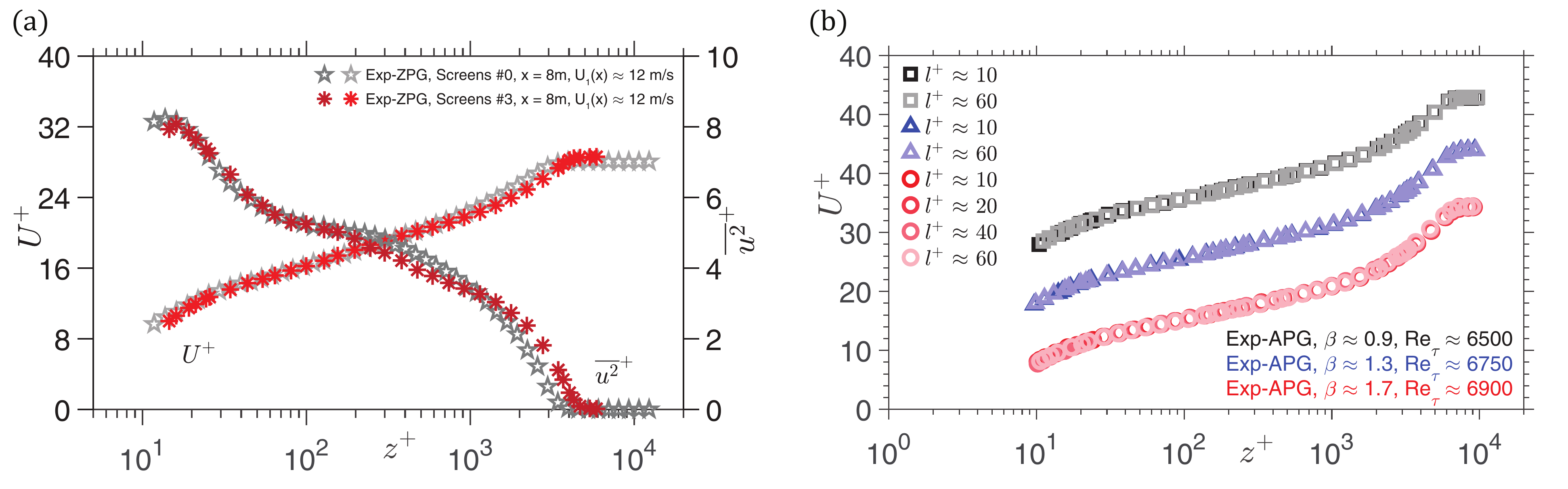}
    \caption{(a) $U^+$ and ${\overline{u^2}}^{+}$ profiles measured in the Melbourne wind tunnel at matched $x$-locations and free-stream velocity conditions for two different outflow screen configurations, Screen\#0 and \#3, as described in $\S$\ref{sec:2setup}.
    (b) Mean streamwise velocity profiles measured for high $Re_{\tau}$ Exp-APG datasets with hotwires of various spatial-resolutions ($l^+$) at various $\beta$, 0.9 $\lesssim$ $\beta$ $\lesssim$ 1.7.
    Symbols and colour scheme for each dataset are as documented in table \ref{tab1}. Data is separated by a shift in the $U^+$-axis for clarity.}
    \label{fig:A1}
\end{figure}

The high $Re_{\tau}$ Exp-APG dataset presented in this study is unique in the sense that the APG TBL develops from a well-established ZPG TBL at $Re_{\tau}$ $\sim$ 4000, as its initial condition ($\S$\ref{sec:2setup}).
This is realized by controlling the outflow from the air bleeds for $x$ $<$ 9\:$m$ of the test section, with outflow screens installed at the test section outlet (\#1,\#2,\#3; figure \ref{fig:2_1tunnelschematic}).
While the near constant $C_P$ distribution in figure \ref{fig:2_1tunnelschematic}(b) does indicate development of a canonical ZPG for $x$ $<$ 9\:$m$  \citep{clauser1956turbulent}, here we reaffirm this claim by comparing streamwise velocity statistics measured at $x$ $\approx$ 8\:$m$ for Screen \#3 configuration, with those measured at matched conditions for Screen \#0 configuration (figure \ref{fig:A1}a).
Here, the data for Screen \#0 configuration is from the published dataset of \citet{hutchins2009}, when there was no control on the air bleeds on the ceiling.
The reasonable match of $U^+$and ${\overline{u^2}^+}$ profiles, obtained from the two different screen configurations, suggests that the TBL developing in the first $\sim$9\:$m$ (for Screen \#3 configuration) can be deemed as near canonical.
This confirms that the APG TBL growing for $x$ $>$ 9\:$m$ has a near-`canonical' upstream-history condition.

\section*{Appendix 2: Mean velocity profiles for Exp-APG with varying spatial-resolutions}\label{app:2MeanVPs}

It is already known \citep{hutchins2009} that the mean velocity profiles ($U^+$) in a ZPG TBL are not influenced by changes in hotwire spatial-resolution ($l^+$) used for measurement.
To confirm that the same is valid for APG TBL, figure \ref{fig:A1}(b) compares $U^+$ profiles obtained at matched $\beta$ and $Re_{\tau}$ but for varying hotwire $l^+$.
Indeed, $U^+$ profiles collapse over one another, suggesting $l^+$ $\lesssim$ 60 doesn't influence the mean velocity profiles, at least in the moderate $\beta$ ($\lesssim$ 2) range considered.

\section*{Appendix 3: Energy decomposition into small and large scale contributions}\label{app:3smallvlarge}

\begin{figure}[t]
\centering
\includegraphics[width=1.0\textwidth]{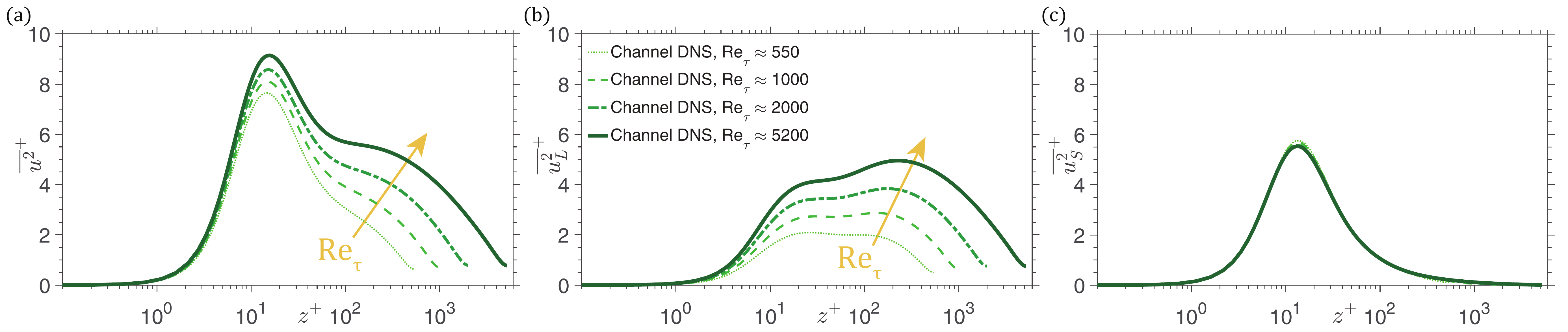}
\caption{(a) ${\overline{u^2}}^+$ decomposed into (b) large scale (${\overline{u^2_{L}}}^+$) and (c) small scale contributions (${\overline{u^2_{S}}}^+$) for DNS dataset of a turbulence channel flow \citep{mklee2015} at various $Re_{\tau}$. Energy decomposition is based on using ${\lambda}^+_{y,c}$ = 300, same as in figure \ref{fig:4_2decomposition}.}
\label{fig:A3}
\end{figure}

This appendix section is presented to reaffirm that the choice of ${\lambda}^+_{y,c}$ = 300 is reasonable to decompose the streamwise normal stress (${\overline{u^2}}^+$) into small (${\overline{u^2_{S}}}^+$; ${\lambda}^+_{y}$ $\le$ ${\lambda}^+_{y,c}$) and large scale contributions (${\overline{u^2_{L}}}^+$; ${\lambda}^+_{y}$ $>$ ${\lambda}^+_{y,c}$).
To this end, figure \ref{fig:A3} shows $\overline{u^2}^+$ decomposed into ${\overline{u^2_{L}}}^+$ and ${\overline{u^2_{S}}}^+$ for the DNS dataset of a turbulent channel flow \citep{mklee2015}.
Consistent with the previous knowledge on small scale behaviour \citep{hutchins2009}, ${\overline{u^2_{S}}}^+$ demonstrates convincing collapse across a decade of $Re_{\tau}$, 550 $<$ $Re_{\tau}$ $<$ 5200, while ${\overline{u^2_{L}}}^+$ increases with $Re_{\tau}$.
Hence, ${\lambda}^{+}_{y,c}$ = 300 seems a reasonable choice \citep{mklee2019} to compare $Re_{\tau}$-trends for ${\overline{u^2_{S}}}^+$ and ${\overline{u^2_{L}}}^+$ between ZPG and APG TBLs.

\section*{Appendix 4: Influence of upstream-history effects on small scale energy}\label{app:4UpstreamHistory}

\begin{figure}[t]
\centering
\includegraphics[width=1.0\textwidth]{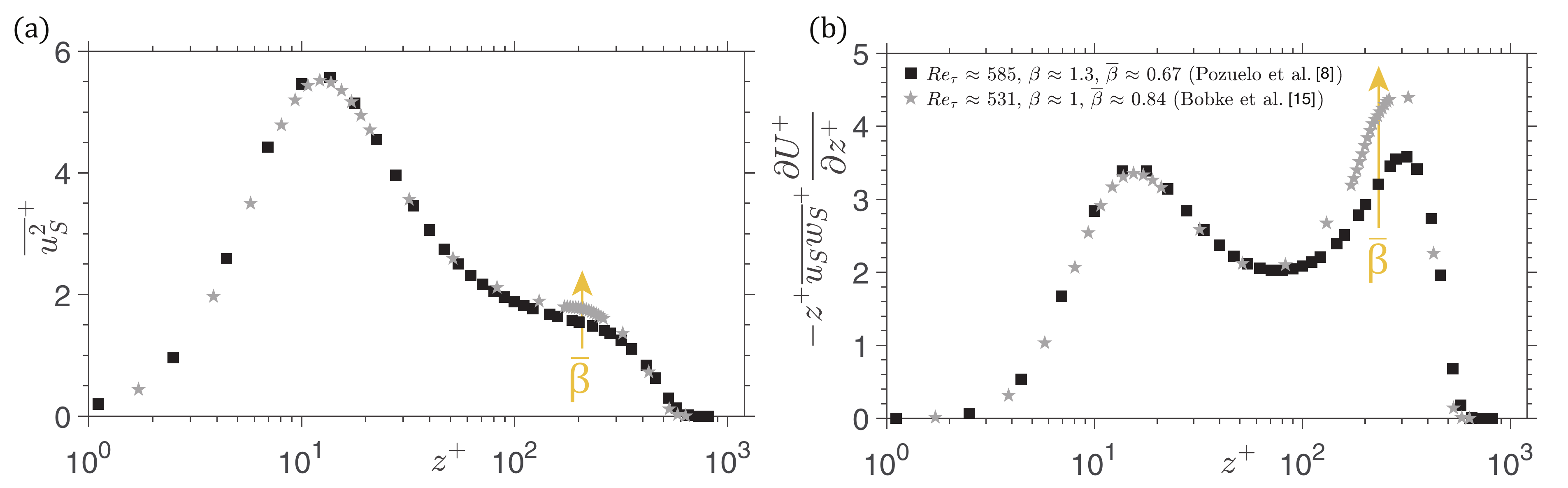}
\caption{small scale contribution to (a) streamwise normal stress (${\overline{u^2_{S}}}^+$) and (b) production of turbulent kinetic energy ($-{{\overline{{u_{S}}{w_{S}}}}^+}{\frac{{\partial}U^+}{{\partial}z^+}}$) compared for LES dataset of \citet{bobke2017history} and \citet{pozuelo2022}.
The comparison is made at similar $Re_{\tau}$ and $\beta$ but different $\overline{\beta}$, to understand the influence of upstream-history. 
Spectral decomposition is based on ${\lambda}^{+}_{y,c}$ $=$ 300.
Estimates for the simulation datasets correspond to their original (\textit{i.e.} well-resolved) grid resolution (${\Delta}{y^+}$, table \ref{tab2})}.
\label{fig:A4}
\end{figure}

This appendix section is presented to reaffirm that the reduction in small scale energy contributions to streamwise normal stress (figure \ref{fig:4_2decomposition}d) and production of turbulent kinetic energy (figure \ref{fig:4_3production}f), discussed in $\S$\ref{sec:4Re_effect}, is not an artefact of upstream-history effects ($\overline{\beta}$).
To this end, figures \ref{fig:A4}(a,b) compare ${\overline{u^2_{S}}}^+$ and $-{{\overline{{u_{S}}{w_{S}}}}^+}{{\partial}U^+ / {\partial}z^+}$ estimated from published LES dataset of \citet{bobke2017history} and \citet{pozuelo2022}, at similar $Re_{\tau}$ and $\beta$ but different $\overline{\beta}$. 
It is evident from the comparison that both ${\overline{u^2_{S}}}^+$ and $-{{\overline{{u_{S}}{w_{S}}}}^+}{{\partial}U^+ / {\partial}z^+}$ increase in the outer region with increasing $\overline{\beta}$.
This confirms the trend documented in table \ref{tab0}, based on our understanding of the literature \citep{tanarro2020}.
Also, this trend is opposite to what is noted on comparing various cases of LES-APG (table \ref{tab2}), suggesting variation in upstream-history effects is weak.
This confirms that the reduction in small scale energy in the outer region of an APG TBL is purely a $Re_{\tau}$-effect.

\begin{acknowledgements}
The authors are grateful to Dr. R. Pozuelo for sharing the LES datasets and acknowledge Mr. P. Manovski for helpful comments/suggestions on the manuscript.
Financial support from the Australian Research Council is gratefully acknowledged. 
R.D. also acknowledges partial financial support by the University of Melbourne through the Melbourne Postdoctoral Fellowship.
R.V. acknowledges the financial support from ERC grant no. `2021-CoG-101043998, DEEPCONTROL'. 
Views and opinions expressed are however those of the author(s) only and do not necessarily reflect those of the European Union or the European Research Council. 
Neither the European Union nor the granting authority can be held responsible for them.
\end{acknowledgements}

\section*{Conflict of interest}
The authors declare that they have no conflict of interest.

\section*{Author contributions}
\textbf{Rahul Deshpande:} Writing – original draft, Conceptualization, Supervision, Methodology, Validation, Investigation, Formal analysis.\\
\textbf{Aron van den Bogaard:} Writing – original draft, Experiments, Investigation, Formal analysis, Validation.\\
\textbf{Ricardo Vinuesa:} Conceptualization, Methodology, Writing – review \& editing, Resources, Investigation.\\
\textbf{Luka Lindić:} Writing – review \& editing, Experimental set up.\\
\textbf{Ivan Marusic:} Supervision, Conceptualization, Writing – review \& editing, Funding acquisition.

%


\end{document}